\documentclass[a4paper,12pt]{article}
\pdfoutput=1
\usepackage{jcappub} % for details on the use of the package, please see the JINST-author-manual
\usepackage{newtxtext,newtxmath}
\usepackage{color,xcolor}
\usepackage[T1]{fontenc}
\usepackage{graphicx}	% Including figure files
\usepackage{latexsym,amsmath,amssymb}
\usepackage{multicol}
\usepackage{tikz}   % Flow chart
\usetikzlibrary{shapes.geometric, arrows}
\usetikzlibrary{arrows}
\usepackage{subfigure,float}
\usepackage{dblfloatfix}
\usepackage{newtxmath}
\usepackage{subcaption}
\usepackage[export]{adjustbox}
\usetikzlibrary{shapes.geometric, arrows}
\graphicspath{{figures/}} % Directory in which figures are stored
\usepackage{multicol}
\usepackage[normalem]{ulem}
\usepackage{parskip} %% <-- added
\usepackage{soul,ulem}
\usepackage{array}
\newcolumntype{P}[1]{>{\centering\arraybackslash}p{#1}}
\def \HI{{\sc Hi}}

\tikzstyle{startstop} = [rectangle, rounded corners, minimum width=3cm, minimum height=1cm, text centered, draw=black, fill=red!30]
\tikzstyle{process} = [rectangle, minimum width=3cm, minimum height=1cm, text centered, draw=black, fill=blue!30]
\tikzstyle{arrow} = [thick,->,>=stealth]
\tikzstyle{decision} = [diamond, 
minimum width=3cm, 
minimum height=1cm, 
text centered, 
draw=black, 
fill=green!30]
\tikzstyle{arrow} = [thick,->,>=stealth]

\arxivnumber{2510.25886v2} % Only if you have one
\title{\boldmath Mitigating gain calibration errors from EoR observations with SKA1-Low AA*}

\author[a,1]{Eeshan Beohar,\note{Corresponding author.}}\emailAdd{eeshan.beohar@gmail.com}
\author[a]{Abhirup Datta,}\emailAdd{abhirup.datta@iiti.ac.in}
\author[a]{Anshuman Tripathi,}\emailAdd{anshumantripathi85@gmail.com}
\author[a]{Samit Kumar Pal,}\emailAdd{palsamitkumar@gmail.com}
\author[a]{and Rashmi Sagar}\emailAdd{rashmisagar777@gmail.com}
\affiliation[a]{Department of Astronomy, Astrophysics \& Space Engineering, Indian Institute of Technology Indore, Indore 453552, India}

\abstract{The observations of the redshifted 21-cm signal from neutral hydrogen are a promising probe for understanding the Cosmic Dawn and the Epoch of Reionisation (EoR). One of the primary obstacles to the statistical detection of the Cosmological signal is the presence of residual foreground arising from gain calibration errors. Previous studies have shown that gain calibration errors as small as 0.01\% can lead to a biased interpretation of the observed signal power spectrum estimation, by nearly an order of magnitude. A recent study further highlights that to accurately retrieve astrophysical parameters, the threshold gain calibration error should be below 0.01\%. This work investigates the impact of residual extragalactic foregrounds arising from gain calibration errors on the efficacy of foreground mitigation strategies. We use an end-to-end pipeline \textsc{21cmE2E} to simulate a realistic sky model and telescope configuration within the 138-146 MHz frequency range and perform a detailed power spectrum analysis across several threshold levels of the gain calibration error. We introduce a hybrid mitigation technique that combines the foreground removal techniques, Gaussian process regression and principal component analysis, with foreground avoidance. Our results indicate that recovery of the \HI\ signal within 2$\sigma$ is possible for calibration gain error of $\leq 1\%$ with minimal loss of power spectrum sensitivity over the scale range $0.05 \leq k \leq 0.5$ Mpc$^{-1}$. We find that gain calibration errors beyond this threshold lead to signal suppression on large scales due to the loss of spectral smoothness of the residual foreground. In effect, this work offers a comparative assessment of three foreground mitigation strategies, removal, avoidance, and a hybrid approach, in the context of future SKA1-Low AA* observations.}

\begin{document}
\maketitle
\flushbottom
\section{Introduction}
\label{sec:intro}
The evolution of the Universe from the post-recombination era to the period following the reionisation of the neutral Intergalactic Medium (IGM) remains largely unconstrained by direct observations. Theoretical models suggest that the first cosmic structures, comprising the earliest stars and galaxies, formed during this interval, corresponding to redshifts between 30 and 7 \cite{barkana_loeb_ionisation_2001}. The emergence of the first stars, a period referred to as the Cosmic Dawn (CD), was driven by small-scale matter density fluctuations induced by gravitational instabilities. Ultraviolet (UV) radiation from these primordial objects gradually ionised the surrounding IGM, initiating the Epoch of Reionisation (EoR) \cite{oh_2006,21cm_in_21cent}.

The neutral hydrogen (\HI) signal from the EoR is expected to be observed at redshifted frequencies in the 50–200 MHz range. Currently, two distinct approaches of observations are employed to detect this signal. Global signal experiments such as EDGES \cite{EDGES3}, REACH \cite{REACH_2022}, RHINO \cite{RHINO_bull}, and SARAS \cite{SARAS2024}, which consist of single radiometer antennas, aim to detect fluctuations in the sky brightness temperature across a broad frequency range. Although EDGES reported a spectral feature that is expected from the absorption of cosmic microwave background (CMB) photons by the neutral hydrogen gas \cite{2018Natur.555...67B}, subsequent analyses, such as those by SARAS \cite{2022NatAs...6..607S}, suggest that the best-fit signal may be of non-astrophysical origin. On the other hand, interferometric arrays such as LOFAR\footnote{Low Frequency Array, \url{http://www.lofar.org}} \cite{LOFAR_2013}, MWA\footnote{Murchison Widefield Array, \url{http://www.mwatelescope.org}} \cite{MWA_2013}, HERA\footnote{Hydrogen Epoch of Reionization Array, \url{http://reionization.org}} \cite{HERA_2017}, uGMRT\footnote{Giant Metrewave Radio Telescope, \url{https://www.gmrt.ncra.tifr.res.in/}} \cite{ygupta_ugmrt}, and the forthcoming SKA\footnote{Square Kilometre Array, \url{https://www.skao.int/en}} \cite{SKA_2015} aim to detect spatial fluctuations in differential brightness temperature statistically. Although these experiments have placed upper limits on the EoR power spectrum \cite{gmrt_UL_21,HERA_UL_23,LOFAR_UL_2025,MWA_UL_23}, a definitive detection of the \HI\ signal remains elusive.

One of the major observational challenges to the detection of the \HI\ signal is bright astrophysical foregrounds, primarily galactic diffuse synchrotron and free-free emissions, along with extragalactic radio sources such as star-forming galaxies and Active Galactic Nuclei (AGN) \cite{smooth_fore_Bharadwaj_2005,smooth_fore_Jeli__2010,smooth_fore_Zahn_2011,arnab_1,arnab_2}. These foregrounds are expected to be 4–5 orders of magnitude brighter than the \HI\ signal. However, the foregrounds typically follow a power-law spectrum, leading to inherent spectral smoothness. In contrast, the \HI\ signal exhibits spectral structures since each frequency corresponds to a different redshift along the line-of-sight (LoS) distances. These contrasting spectral properties form the basis for foreground mitigation strategies.

The effectiveness of foreground mitigation is typically evaluated using cylindrically averaged two-dimensional power spectra (2D PS), which is a function of $k$-modes perpendicular to ($k_{\perp}$) and along ($k_{\parallel}$) the LoS (see section \ref{2D_PS}). The limits on $k_{\perp}$ and $k_{\parallel}$ are governed by the baseline distribution of the array and spectral bandwidth/resolution, respectively. The 2D PS contains distinct regions where the EoR signal, astrophysical foregrounds, and instrumental systematics dominate. Spectrally smooth foregrounds are expected to dominate low $k_\parallel$ modes, while the \HI\ signal is strongest on large scales (low $k_\perp$). Dominated only by thermal noise, the \HI\ signal remains uncontaminated towards higher $k_\parallel$ values, forming the so-called ``EoR window'' \cite{eor_window_math_form}. However, instrument chromaticity leads to mode-mixing, where foreground power leaks into the EoR window, forming the ``foreground wedge'' \cite{Datta_et_al,Morales_2012_wedge,Trott_2012}.

Foreground mitigation approaches are broadly classified into foreground removal and foreground avoidance \cite{Datta_et_al,chapman_fore_review}. Foreground avoidance is an approach that involves discarding the contaminated modes inside the foreground wedge, probing the accessible EoR window. On the other hand, foreground removal techniques aim to fit a foreground model and subtract it, preserving power spectrum sensitivity across all scales. Although both of these approaches have contributed to current upper limits, each has tradeoffs. Foreground avoidance is robust to systematics but sacrifices sensitivity, especially at large scales, because it discards low $k_\parallel$ modes \cite{pober_fore_avoid_2014ApJ}. Conversely, non-parametric subtraction techniques \cite{FASTICA_12,GMCA_13,CCA_14,ghosh_lofar,blind_fore_sub_alonso,mertens_gpr_18}, which rely on the frequency-frequency covariance of statistically independent components, are susceptible to signal loss due to leakage of the cosmological signal into the foreground model \cite{Hothi_21}. The inaccuracies in the foreground model are enhanced by systematics such as calibration errors, incomplete sky models \cite{cal_barry_16,cal_dillon_18,cal_trott_16,cal_wise_17,aishrila_gain_err,Datta_gain_err,joseph_incom_sky_model_18}, beam uncertainties \cite{chokshi_beam_uncertainities_24,kim_beam_2022,nasir_beam}, mutual coupling \cite{cali_gain_err_mutual_coup,kern_mutual,eloy_spectr_17}, and ionospheric distortions \cite{samit_ionosphere,Iono_MWA_2022}. These effects can introduce excess variance or mimic the spectral characteristics of the \HI\ signal, leading to false detections.

In this work, we focus on mitigating the impact of gain calibration errors using post-calibration residual foreground mitigation techniques. We use an end-to-end simulations pipeline to assess the efficacy of Principal Component Analysis (PCA) \cite{blind_fore_sub_alonso} and Gaussian Process Regression (GPR) \cite{mertens_gpr_18} in recovering the 21\,cm power spectrum under varying gain error conditions. In addition to purely foreground avoidance or subtraction, inspired by 'hybrid' foreground mitigation techniques that suggest avoidance along with subtraction \cite{chapman_effectforegroundmitigationstrategy,hybrid_paper_2018,epand_eor_liu_2014}, we apply PCA and GPR to suppress foreground power at low $k_{\parallel}$ sufficiently to open the EoR window for avoidance-based analysis. We demonstrate that this hybrid approach allows partial recovery of large-scale $k$-modes in the 2D PS, where the \HI\ signal is strongest, and hence improves the signal-to-noise ratio (SNR). Our primary objective is to mitigate the residual foreground resulting from gain calibration errors as much as possible, while maintaining power spectrum sensitivity.

This paper is structured as follows: Section~\ref{sec:Simulations} describes the simulation setup, including both ideal and gain-error–induced SKA1-Low AA* observations. Section~\ref{sec:FG_techniques_PS} outlines the residual foreground mitigation techniques and their implementation. In Section~\ref{sec:results}, we present the results of the 1D and 2D PS under various calibration conditions. Finally, Section~\ref{sec:conclusion} discusses our conclusions and outlines future directions. Throughout the work, we used the best-fitting cosmological parameters from the Planck 2018 results \cite{Planck18}.

\section{Simulations}
\label{sec:Simulations}
In this section, we present an end-to-end framework that simulates synthetic SKA-Low AA* data, introduces residual gain calibration errors, and produces corrupted visibilities as a result.

\subsection{Synthetic observations}
\label{subsec:Synth_Obs}
\begin{figure}
    \centering
\begin{tikzpicture}[
    node distance=1.75cm,
    every node/.style={align=center},
    startstop/.style={rectangle, rounded corners, draw, minimum width=3cm, minimum height=1cm},
    process/.style={rectangle,rounded corners, draw, minimum width=3cm, minimum height=1cm},
    arrow/.style={thick,->}
]

% Inputs
\node (fg)  [process] {Foregrounds (FG)\\ Point Sources};
\node (sig) [process,left of=fg, xshift=-1.75cm] {\HI\ Signal\\ (21cmFAST)};
% Merge / common step
\node (layout) [process,right of=fg, xshift=1.75cm] {Telescope Layout\\SKA1-LOW AA*};
% Merge / common step
\node (obs_param) [process, below of=fg] {OSKAR Simulations};
% Merge / common step
\node (obs_vis) [process, below of=obs_param] {Model Visibility};
% Sub-processes
\node (gain_err) [process,below of=obs_vis] {Calibration Gain error};
\node (res_vis) [process,below of=gain_err] {Residual Visibility\\(Model-Corrupted)};
\node (fg_mitigation) [process,below of=res_vis] {Foreground Mitigation};

\node (fg_sub) [process,below of=fg_mitigation] {Subtraction\\ (PCA,GPR)};
\node (fg_avoid) [process,left of=fg_sub,xshift=-2.5cm] {Avoidance\\ (horizon or horizon + buffer)};
\node (fg_hybrid) [process,right of=fg_sub, xshift=2.5cm] {Hybrid Methods\\(subtraction+avoidance)};

\node (PS) [process,below of=fg_sub] {Power Spectrum\\1D,2D PS};

% Arrows
\draw [arrow] (sig) -- (obs_param);
\draw [arrow] (fg) -- (obs_param);
\draw [arrow] (layout) -- (obs_param);

\draw [arrow] (obs_param) -- (obs_vis);
\draw [arrow] (obs_vis) -- (gain_err);
\draw [arrow] (gain_err) -- (res_vis);
\draw [arrow] (res_vis) -- (fg_mitigation);

\draw [arrow] (fg_mitigation) -- (fg_sub);
\draw [arrow] (fg_mitigation) -- (fg_avoid);
\draw [arrow] (fg_mitigation) -- (fg_hybrid);

\draw [arrow] (fg_sub) -- (PS);
\draw [arrow] (fg_avoid) -- (PS);
\draw [arrow] (fg_hybrid) -- (PS);

\end{tikzpicture}
\caption{A flowchart depicting the workflow of \textsc{21cmE2E}, along with the residual foreground mitigation techniques and power spectra analysis done in this work.}
    \label{fig:flowchart}
\end{figure}
We used a 21-cm end-to-end (\textsc{21cmE2E}) pipeline \cite{aishrila_gain_err, samit_ionosphere, 2023Aishrila} to perform synthetic observations for the SKA-Low telescope layout AA*. This pipeline employs the \textsc{OSKAR}\footnote{\url{https://github.com/OxfordSKA/OSKAR}} software package \cite{OSKAR} for simulating visibilities for SKA1-Low configurations. The resulting measurement sets were then processed using the Common Astronomy Software Applications (CASA)~\citep{casa} package for reading, calibration, and further analysis of the visibility data.
In this simulated observation, the sky was observed with a phase-center at $\alpha$ = 15h00m00s and $\delta$ = -30$^\circ$ for a duration of 4 hours ($\pm$ 2 HA). The observing bandwidth of the lightcone spans 8 MHz with a channel separation of 125 kHz. 
A flow chart showing the workflow of \textsc{21cmE2E} pipeline integrated with foreground mitigation techniques is shown in the Figure~\ref{fig:flowchart}. The observation parameters used in the simulations are listed in the Table~$\ref{tab:obs_params}$. In the subsections, we describe the realistic sky models and telescope configuration used to perform the simulation with the \textsc{21cmE2E} pipeline.
\begin{table}
\centering
    \begin{tabular}{c|c}
        \hline
        Parameters & Value\\
        \hline
        Central Frequency (MHz) & 142 (z$\sim$9) \\
        Bandwidth (MHz) & 8 \\
        Number of Channels & 64 \\
        Field of View & 3$^\circ$\\
        Number of stations & 231 \\
        Maximum Baseline (km) & 3.15 \\
        Duration of observation (hrs) & 4 \\
        Time resolution (sec) & 120 \\
        PSF FWHM & 0.38 Mpc$^{-1}$ \\
        \hline
    \end{tabular}
    \caption{Observation parameters used for SKA1-Low AA* simulations.}
    \label{tab:obs_params}
    \end{table}    
\subsubsection{Telescope model}
\label{subsubsec:telescope_model}

\begin{figure}
    \centering
    \includegraphics[width=.45\textwidth]{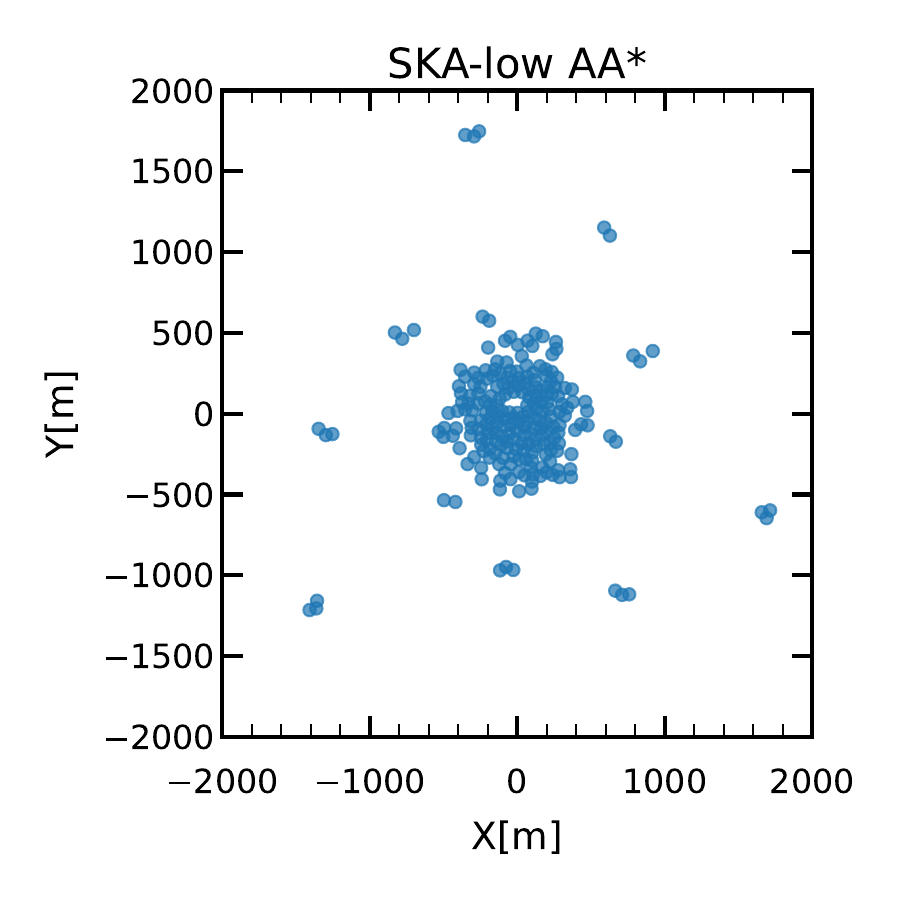}
    \quad
    \includegraphics[width=.45\textwidth]{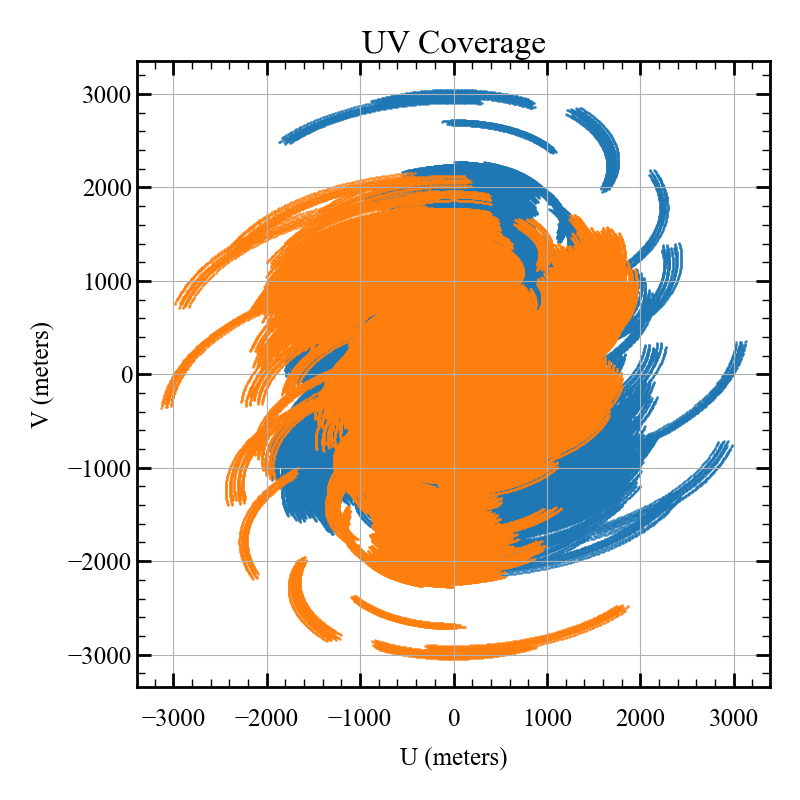}
    \caption{\texttt{Left} : The SKA1-Low AA* stations layout for central 2000 m cutout. \texttt{Right}: UV coverage for the observational parameters in the Table \ref{tab:obs_params}, with integration time of 4 hours ($\pm 2$\,HA). The orange and blue colours represent the (U, V) and (-U,-V) points, respectively.}
    \label{fig_1:ska_low_layout_UVcov}
\end{figure}
The SKA1-Low AA* \cite{ska_low_aastar_seethapuram_sridhar_2025_16951020} layout (Figure~\ref{fig_1:ska_low_layout_UVcov}) is a defined configuration of the SKA1-Low array, comprising 307 stations (at the time of writing) with a maximum baseline of approximately 73.4\,km \cite{aa*_baseline_design_2013}. It preserves the essential features of the full array design, with a dense central core, in which roughly 50\% of the stations are packed within a 1\,km diameter, and an extended distribution along three logarithmically spaced spiral arms \cite{ska_perf_braun_2019}. In our work, we restrict the array to stations within a 2\,km radius to mitigate foreground leakage on larger scales \cite{Koopmans_2015_stat_detection}, also reducing computational demands. The visibilities sampled (UV coverage) for the SKA1-Low AA* layout are shown in the right plot of Figure~\ref{fig_1:ska_low_layout_UVcov}.
\begin{figure}
  \centering
    \includegraphics[width=.45\textwidth]{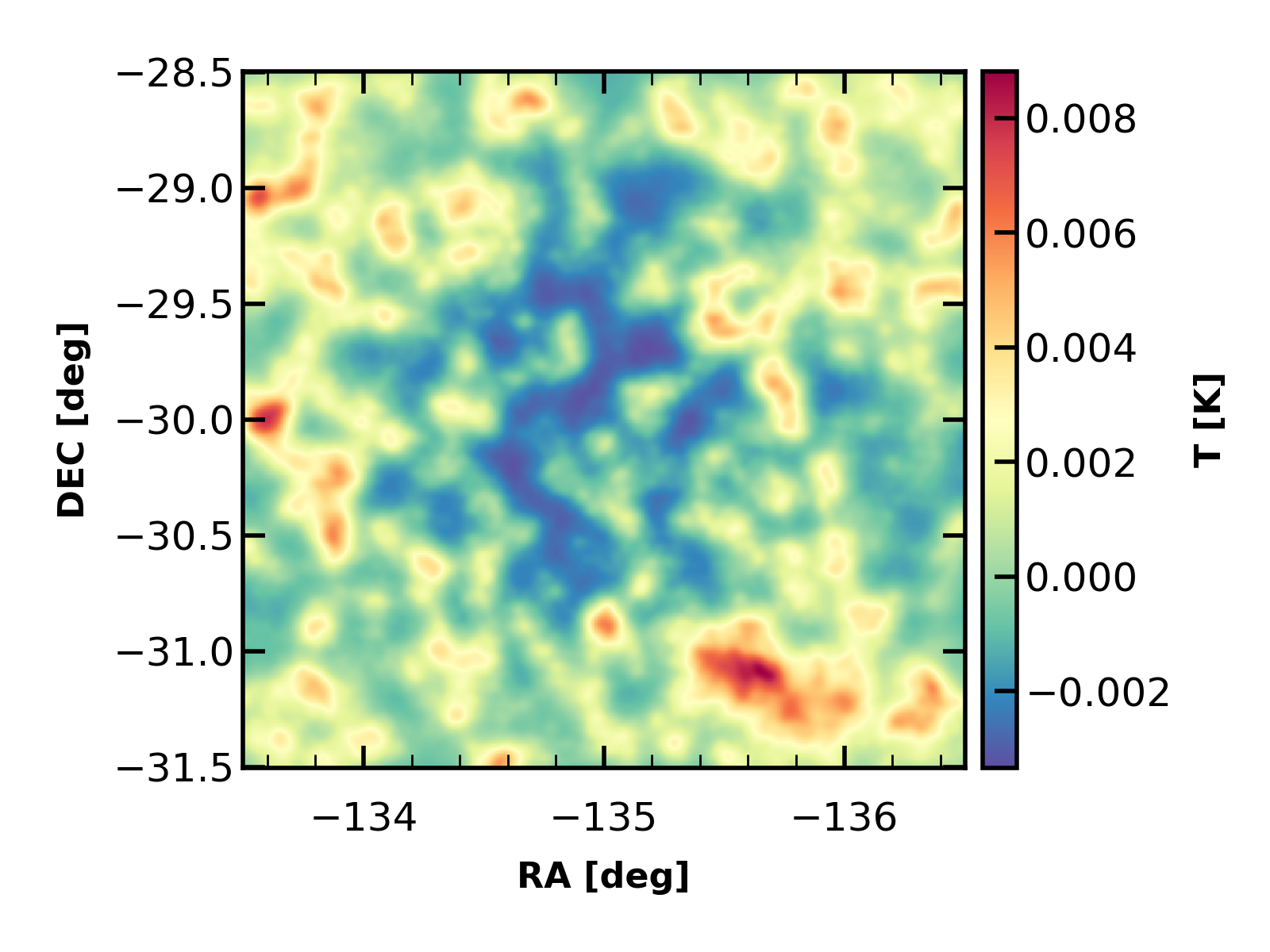}
    \includegraphics[width=.45\textwidth]{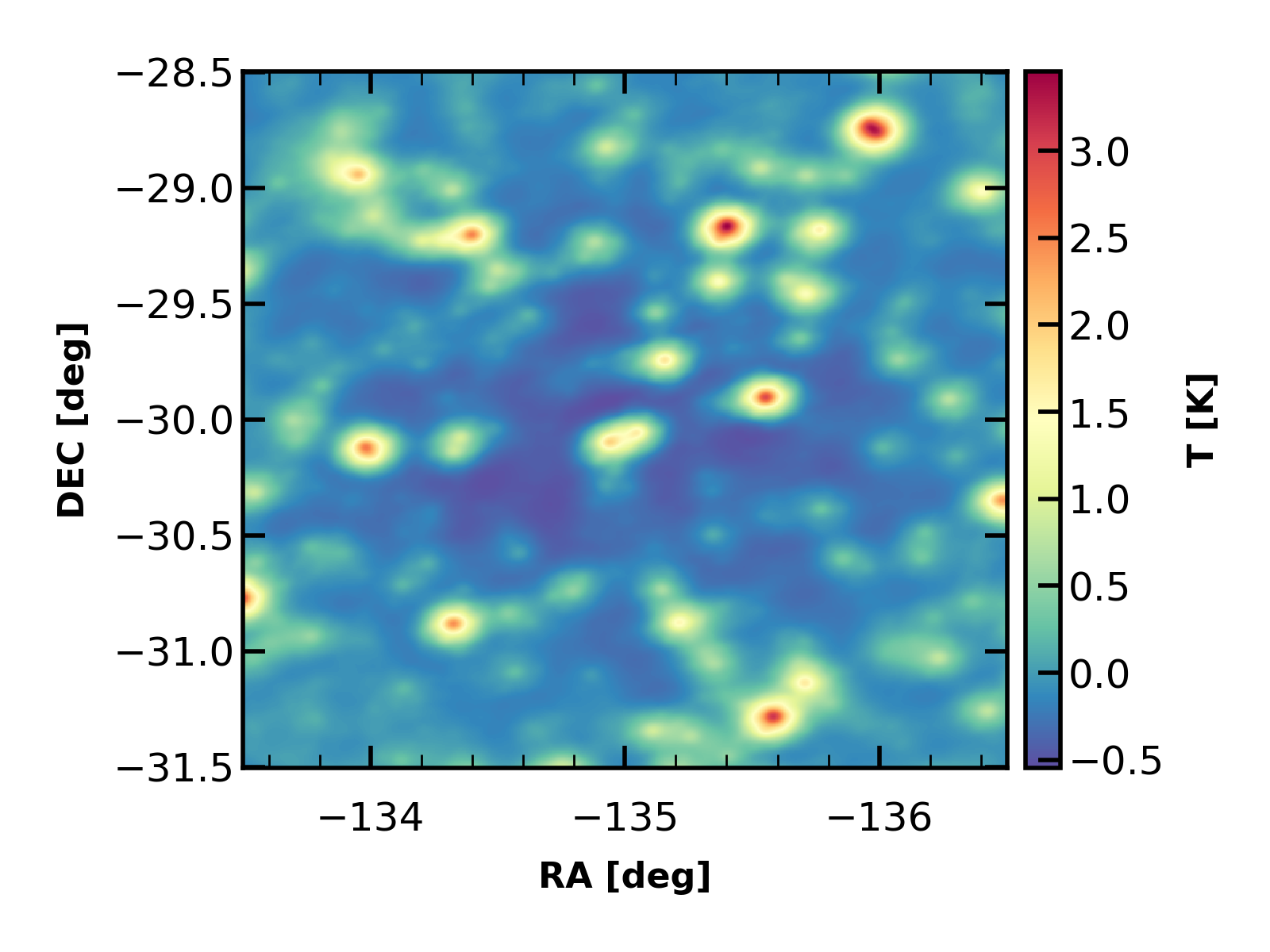}
    \caption{\texttt{Left}: Simulated $3^\circ \times 3^\circ$ slice of the observed \HI\ signal. \texttt{Right}: Residual map after sky-model subtraction with 1\% gain calibration error at 144.25 MHz.}
    \label{fig_2:gain_err_sig_image_comp}
\end{figure}

\subsubsection{EoR Signal}
\label{subsubsec:HI}
For \HI\ simulations, we use the semi-numerical EoR simulation, \textsc{21cmFAST}\cite{Mesinger_2010,Murray2020}, to generate the brightness temperature map ($\delta T_b$). We consider a spatial co-moving volume of $500\times500\,$h$^{-1}$Mpc$^{2}$, with pixel size of $232\times232$, a redshift range of 7.4 to 8.2  and channel width of 125\,kHz respectively. This work used the existing simulated \HI\ lightcone from Tripathi et al. \cite{tripathi_calibration}. Since we do not include thermal noise in our data, we consider this to be the ground truth for all power spectrum results. 
\subsubsection{Point source model}
\label{subsubsec:fg}
The foreground model used in this study consists exclusively of compact sources extracted from the Tiered Radio Extragalactic Continuum Simulation (T-RECS) catalogue \cite{Bonaldi_2018}. The sources were redistributed around a declination of -30$^\circ$ and the flux densities were interpolated to 142\,MHz by assuming a smooth power law relation with spectral index $\alpha$ equal to -0.8. Our sky model consists of 2522 point sources within our simulated FoV of 4$^\circ$. We refer to Section 2 of \cite{samit_ionosphere} for a comprehensive overview of the sky model. It is important to note that the galactic diffuse foreground models are not included in our simulation.

\subsubsection{Gain Calibration Errors}\label{subsec:gain_err}
Gain calibration errors are defined as imperfections in the amplitude and phase of these complex gain terms, which may arise from various instrumental factors, including temperature fluctuations in electronic components, and time-dependent variations in antenna response \cite{cal_barry_16}. For the scope of this work, we have excluded calibration bandpass errors.

The errors due to imperfect source model calibration can introduce spurious features that either mimic or entirely obscure the faint 21-cm cosmological signal \cite{Datta_gain_err}. Simulations of frequency-independent calibration errors by  Morales et al. \cite{morales_wedge_cali_errors_2018} suggest a more prominent contamination in the reconstructed (image-based) power spectrum than in the visibility-based power spectrum, estimates also known as the delay spectrum. Therefore, we expect that the contamination in our estimation also depends on the quality of the instrumental Point Spread Function (PSF) \cite{cal_barry_16}. Mazumder et al.  \cite{aishrila_gain_err} showed that for SKA1-Low, gain calibration errors as low as 0.01\% can significantly contaminate the EoR window, thereby reducing the SNR. The primary goal of this work is to mitigate the effects of these calibration errors during post-processing, treating them as perturbations in the gain solutions that arise from an unoptimised calibration algorithm.

We define the complex gain $g_i$ for the $i^\text{th}$ antenna as:
\begin{equation}
\label{eq:gain_def}
g_i = (a_i + \delta a_i) \exp\left[ -i(\phi_i + \delta \phi_i) \right]
\end{equation}
where $a_i$ and $\delta a_i$ are the amplitude and its error, and $\phi_i$ and $\delta \phi_i$ are the phase and its error, respectively.

In an ideal situation, $a_i = 1$ and $\phi_i = 0$, so we can simplify Equation~\ref{eq:gain_def} as:
\begin{equation}
\label{eq:gain_simplified}
g_i = (1 + \delta a_i) \exp\left[ -i \delta \phi_i \right]
\end{equation}

Assuming the amplitude and phase errors are independent and drawn from a normal distribution, we rewrite:
\begin{equation}
\label{eq:gain_stochastic}
g_i = \left(1 + \mathcal{N}(0,\sigma_\text{err})\right) \exp\left[ -i \mathcal{N}(0,\sigma_\text{err}) \right]
\end{equation}
For each timestamp and throughout the 4-hour observation, these errors are multiplied by the observed visibilities before being subtracted from the sky model. This simulates the percentage inefficiency of the sky-based calibration algorithm and the residual foreground resulting from imperfect subtraction. In this work, we consider three separate cases of a progressively contaminated reconstructed sky varying $\sigma_\text{err}$ to 0.1\%, 1\%, and 10\%, respectively.

\subsubsection{Imaging}
The synthesised maps were generated from the simulated visibilities from the \textsc{21cmE2E} pipeline using \textsc{WSCLEAN} \cite{Offringa_2014_wsclean}. The visibilities were gridded, and each sub-band was imaged independently. We stack each sub-band along the frequency axis, making an image cube ($\theta_x, \theta_y, \nu$). We apply Gaussian tapering and W-projection to produce naturally weighted dirty images of 4$^\circ$ FoV with a pixel size of $15\times15$\, arcsec. Figure~\ref{fig_2:gain_err_sig_image_comp} shows a 3$^\circ$x3$^\circ$ slice of \HI\ map and point source subtracted residual map with 1\% at 144.25 MHz. Although we simulate 4$^\circ$ sky areas around the zenith for all the components of the sky signal, we perform the power spectrum estimation using only 3$^\circ$ from the phase-centre, corresponding to SKA-low's anticipated primary beam \cite{ska_low_braun_2019}. The effect of PSF starts dominating around $k_{\perp}$ $\sim$ 0.38 Mpc$^{-1}$. Out of the spherically averaged $k$-modes available by our observations 0.05 $\leq$ k $\leq$ 1.57 Mpc$^{-1}$, we focus on large scales of cosmological interest and perform power spectrum analysis for 0.05 $\leq$ k $\leq$ 0.5 Mpc$^{-1}$.

\subsection{Power spectrum estimation}
\label{2D_PS}
In this section, we provide an overview of image-based (reconstructed) power spectrum estimation used in this work to evaluate cylindrically and spherically averaged 2D and 1D power spectra, respectively.

Let $T(\nu,\mathbf{r})$ be the observed brightness map, and $\tilde{T}(\tau,\mathbf{u})$ be its normalized discrete Fourier transformation, defined as:
\begin{equation}
\label{eq:ps_1}
\begin{aligned}
    \tilde{T}(\tau,\mathbf{u}) = \frac{10^{-26}c^{2}}{2k_{B}N_{v}\nu^{2}\Omega_{PSF}} \sum_{\nu,\mathbf{r}}T(\nu',\mathbf{r'}) e^{-2\iota \pi(\mathbf{u\cdot r' + \nu' .\tau})}
\end{aligned}
\end{equation}
where $N_{v} = N_x\times N_y \times N_{\nu}$ is the number of voxels in our data cube and corresponding voxel resolution as $\delta_{v} = \delta_x\times \delta_y \times \delta_{\nu}$, where $\delta_{x}$, $\delta_{y}$ and $\delta_{\nu}$ are spatial and spectral resolution, respectively. Before performing the spectral Fourier transform, we divide the temperature voxels with PSF in the $(\nu,\mathbf{u})$ domain and then apply a Blackman-Harris frequency window taper \cite{blackman_harris_window}. Connecting radio interferometry to physical cosmology, we define wavenumber $\mathbf{k}=(k_{\perp},k_{\parallel})$ as:
\begin{equation}
\label{eq:ps_2}
    \begin{aligned}
        k_{\perp} = \frac{2\pi|\mathbf{u}|}{D_{M}(z)},
        \quad
        k_{\parallel} \approx \frac{2\pi\mathbf{\tau}H_{0}E(z)}{c(1+z)^{2}}
    \end{aligned}
\end{equation}
where $D_M(z)$ and $\Delta D$ are conversion factors from angle and frequency to co-moving distance \cite{conv_mcquin,conv_morales_2004}, B is the frequency bandwidth, $H_{0}$ is Hubble constant, $\nu_{\rm 21}$ is the frequency of the 21-cm line, $E(z)$ is the dimensionless Hubble parameter defined as $E(z) = (\Omega_{M}(1+z)^{3}+\Omega_{K}(1+z)^{2} + \Omega_{\Lambda})^{1/2}$ \cite{hogg_1999}, and $z$ is the reference redshift at the centre of the observed bandwidth.\\
We define the cylindrical averaged 2D PS as:
\begin{equation}
\label{eq:ps_3}
    \begin{aligned}
        P(k_{\perp},k_{\parallel}) = \mathbb{V}\langle|T(\tau,\mathbf{u})|^{2}\rangle_{k_\perp,k_\parallel},
        \quad
        \mathbb{V} = \frac{N_{v}\delta_{v}D^{2}_M(z)\Delta D}{\Omega_{PB}B}
    \end{aligned}
\end{equation}
where $\mathbb{V}$ is co-moving volume, $\delta_{x}$, $\delta_{y}$ and $\delta_{\nu}$ are spatial and spectral resolution, respectively.

To quantify the excess residual power in the 2D PS for different cases, we define the absolute fractional error maps as:
\begin{equation}
\label{eq:2dps_frac_err}
\begin{aligned}
\Delta{P}_{i}(k_{\perp},k_{\parallel}) = \frac{|P_{i}(k_{\perp},k_{\parallel}) - P_{21}(k_{\perp},k_{\parallel})|} {P_{21}(k_{\perp},k_{\parallel})}
\end{aligned}
\end{equation}
where `i' corresponds to gain error, PCA-, or GPR-filtered power spectrum.\\
The dimensionless 1D PS for spherically averaged $k$-modes $k = \sqrt{k_{\perp}^{2}+k_{\parallel}^{2}}$ is defined as:
\begin{equation}
\label{eq:1d_ps}
    \begin{aligned}
        P(k) = \frac{k^{3}}{2\pi^{2}}(\mathbb{V}\langle|T(\tau,\mathbf{u})|^{2}\rangle_{k}),
    \end{aligned}
\end{equation}
We define the wedge line as:
\begin{equation}
\label{eq:wedge_def}
\begin{aligned}
    k_{\parallel} \leq k_{\perp} \frac{D_M(z)E(z)H_{0}} {(1+z)c} \sin{\theta}
\end{aligned}
\end{equation}
where the equation converts to an expression for the horizon line for $\theta$ = $\pi/2$.\\
To compare the effects of foreground avoidance and subtraction on power spectrum sensitivity and SNR per $k$-bin, we use \textsc{21cmSense}\footnote{\url{https://github.com/rasg-affiliates/21cmSense}} \cite{21cmSense_Murray2024, pober_2013, pober_fore_avoid_2014ApJ} to evaluate the theoretical thermal noise for 100 hours of SKA1-Low AA* observations. The 1D thermal noise uncertainty is defined as:
\begin{equation}
\label{eq:1d_ps_err_th}
\Delta^{2}_{\rm th}(k) = \frac{k^{3} D^{2}_M(z)\Delta D}{2\pi^{2}} \cdot \frac{\Omega^{2}_P}{\Omega_{\rm PP}} \cdot \frac{T^{2}_{\rm sys}}{2t}
\end{equation}

\begin{figure}
  \centering
  \includegraphics[width=.75\textwidth]{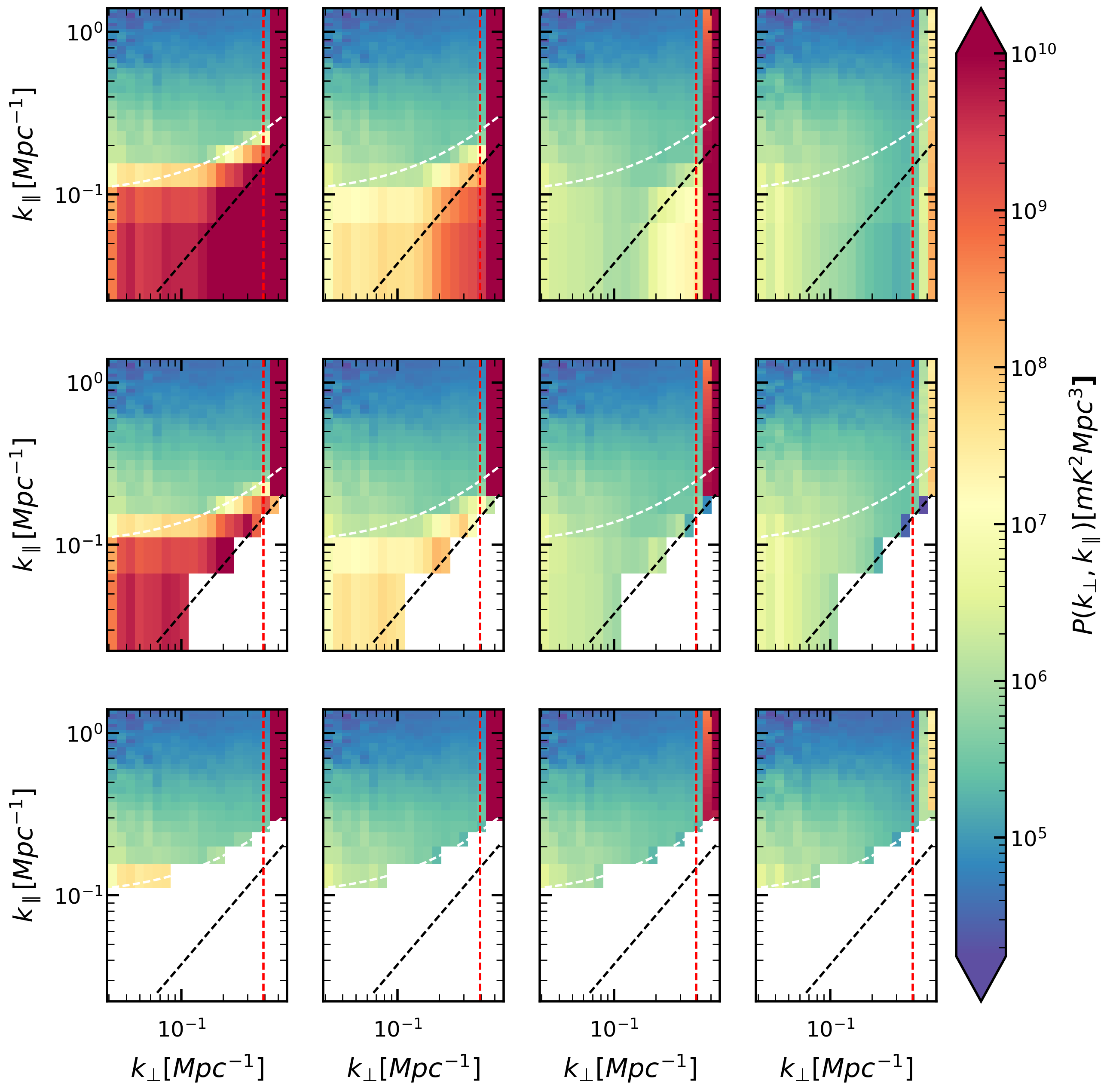}
  \caption{From \texttt{top to bottom}, we show cylindrically averaged 2-D PS for three scenarios: no foreground avoidance, avoidance about the horizon, and avoidance about horizon plus 0.1 Mpc$^{-1}$ buffer, respectively. Proceeding \texttt{right to left}, we show 2-D PS for the true \HI\ (perfect foreground subtraction), and for the point-source subtracted residual data with 0.1\%, 1\%, and 10\% gain calibration error. The minimum and maximum baselines were taken to be $\sim48 \lambda$ and $\sim948 \lambda$. The black and white dotted lines represent the horizon limit and horizon plus 0.1 Mpc$^{-1}$ buffer, respectively. The vertical red dotted line marks the scale at which PSF starts to dominate.}
  \label{fig_3:gain_err_2d_ps_com}
\end{figure}

\begin{figure}
  \centering
  \includegraphics[width=.75\textwidth]{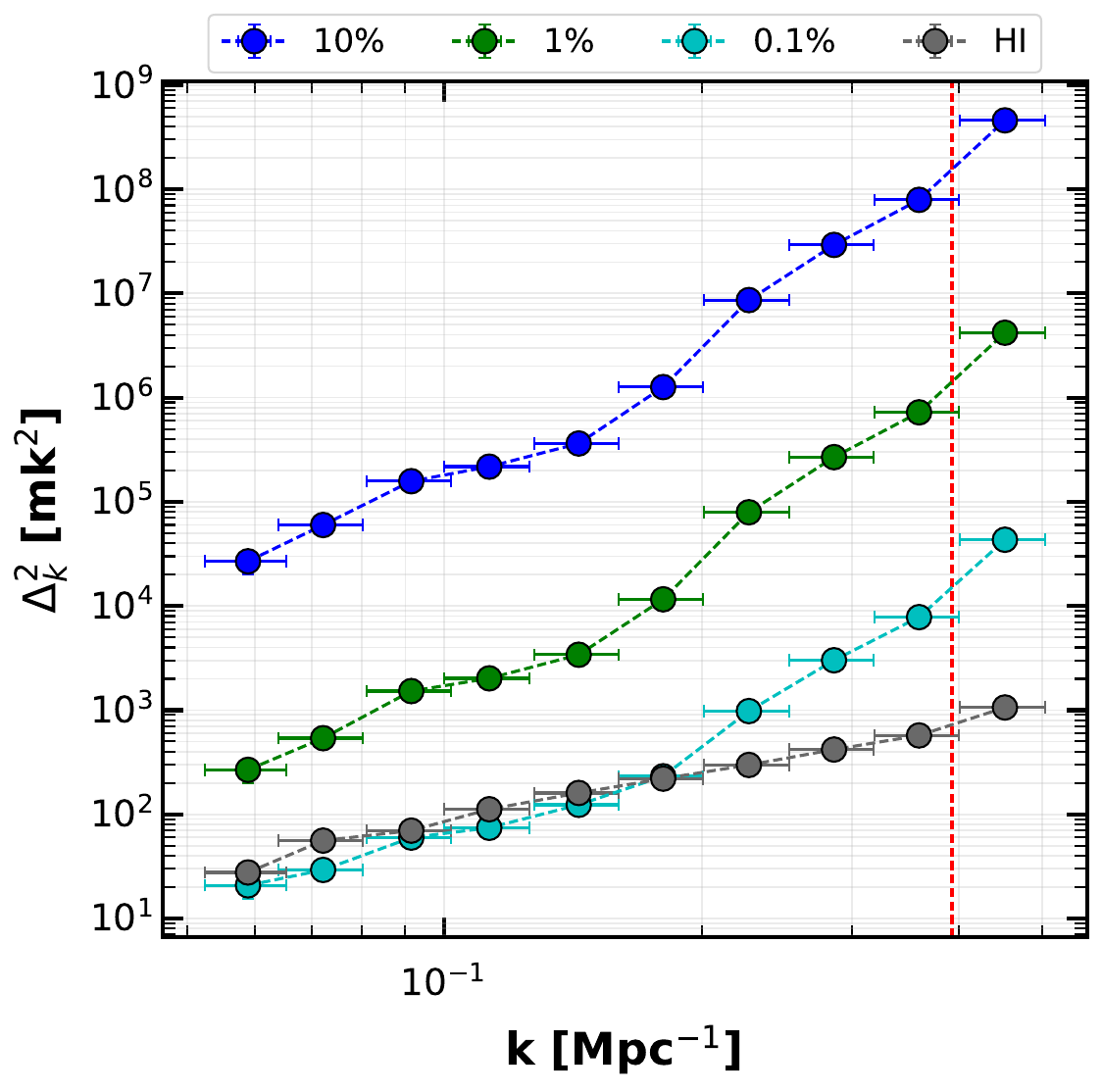}
  \caption{The 1D PS on logarithmically spaced, spherically averaged $k$-bins from 0.05 to 0.5 Mpc$^{-1}$. The markers represent 1-D power spectra for residual visibilities with various contamination percentages of gain calibration errors, as well as the true \HI. The vertical error bars indicate total uncertainty due to 2$\sigma$ sample variance and theoretical thermal noise; the horizontal bars show bin widths. The vertical red dotted line marks the scale at which PSF starts to dominate.}
  \label{fig_4:gain_err_1d_ps_comp}
\end{figure}

The uncertainty due to sample variance is represented as a $2\sigma$ error:
\begin{equation}
\label{eq:1d_ps_err_sv}
\sigma^{2}_{i} = \frac{\Delta^{2}(k \in k_{i})}{N_{i}}
\end{equation}
where $\Delta^{2}(k \in k_{i})$ denotes the spherically averaged power in the $i^{\mathrm{th}}$ $k$-bin, and $N_i$ is the number of independent modes contributing to that bin.

The total uncertainty in each $k$-bin is taken as the sum of the thermal and sample variance contributions from equations~\ref{eq:1d_ps_err_th} and~\ref{eq:1d_ps_err_sv}.

We define the SNR in each $k$-bin as:
\begin{equation}
\mathrm{SNR}(k) = \frac{\Delta_{\rm HI}^2(k)}{\Delta_{\rm N}^2(k)},
\end{equation}
where $\Delta_{\rm HI}^2(k)$ is the \HI\ power spectrum and $\Delta_{N}^2(k)$ is the total uncertainty (sample variance + 100\,hrs of observation of instrumental noise).

The plots in the last column of Figure~\ref{fig_3:gain_err_2d_ps_com}, and the gray markers in Figure~\ref{fig_4:gain_err_1d_ps_comp} show the 2D PS and 1D PS of a realisation of the \HI\ cube as observed with our telescope model.
As can be seen from the 2-D cylindrical power spectrum in Figure~\ref{fig_3:gain_err_2d_ps_com}, residual foreground contamination in the EoR window increases with increasing residual gain calibration error. Specifically, for the calibration inaccuracy of 0.1\%, the excess residual power manifests itself as a mode-mixing feature \cite{mertens_gpr_18} at high $k_\perp$ modes. Thus, foreground avoidance about the horizon line appears viable (see Section~\ref{subsec:results_0.1}). However, for higher corruption levels, the familiar foreground wedge, which includes both intrinsic smooth partitions and mode-mixing features, becomes prominent.

The impact of gain errors is also evident in the spherically averaged 1D PS (Figure~\ref{fig_4:gain_err_1d_ps_comp}) as excess power above the true \HI\ signal. This further confirms the contamination introduced by imperfect calibration.

\section{Residual Foreground Mitigation}
\label{sec:FG_techniques_PS}
In this section, we provide a brief overview of the foreground mitigation techniques employed in this work, with a focus on their application to EoR observations.
\subsection{Foreground Avoidance}
\label{subsec:FG_avoid}
As discussed in Section~\ref{subsec:gain_err}, gain calibration errors increasingly contaminate the EoR window, necessitating a more conservative approach to foreground avoidance. In particular, applying buffers beyond the geometric horizon may be required to exclude contaminated $k$-modes. However, this comes at the cost of reduced sensitivity, particularly on larger scales, due to the loss of low-$k$ modes critical for statistical detection. To investigate this tradeoff, we examine three foreground avoidance scenarios with varying wedge boundaries, similar to the optimistic and pessimistic cases considered in \cite{pober_fore_avoid_2014ApJ}.

In the first scenario, we assume ideal subtraction performance and apply no foreground avoidance, thereby retaining all $k$-modes for power spectrum estimation. In the second scenario, we apply a wedge cut at the horizon limit. In the third, more conservative case, we impose an additional buffer of 0.1 Mpc$^{-1}$ beyond the horizon line.

\begin{figure}
  \centering
  \includegraphics[width=0.75\textwidth]{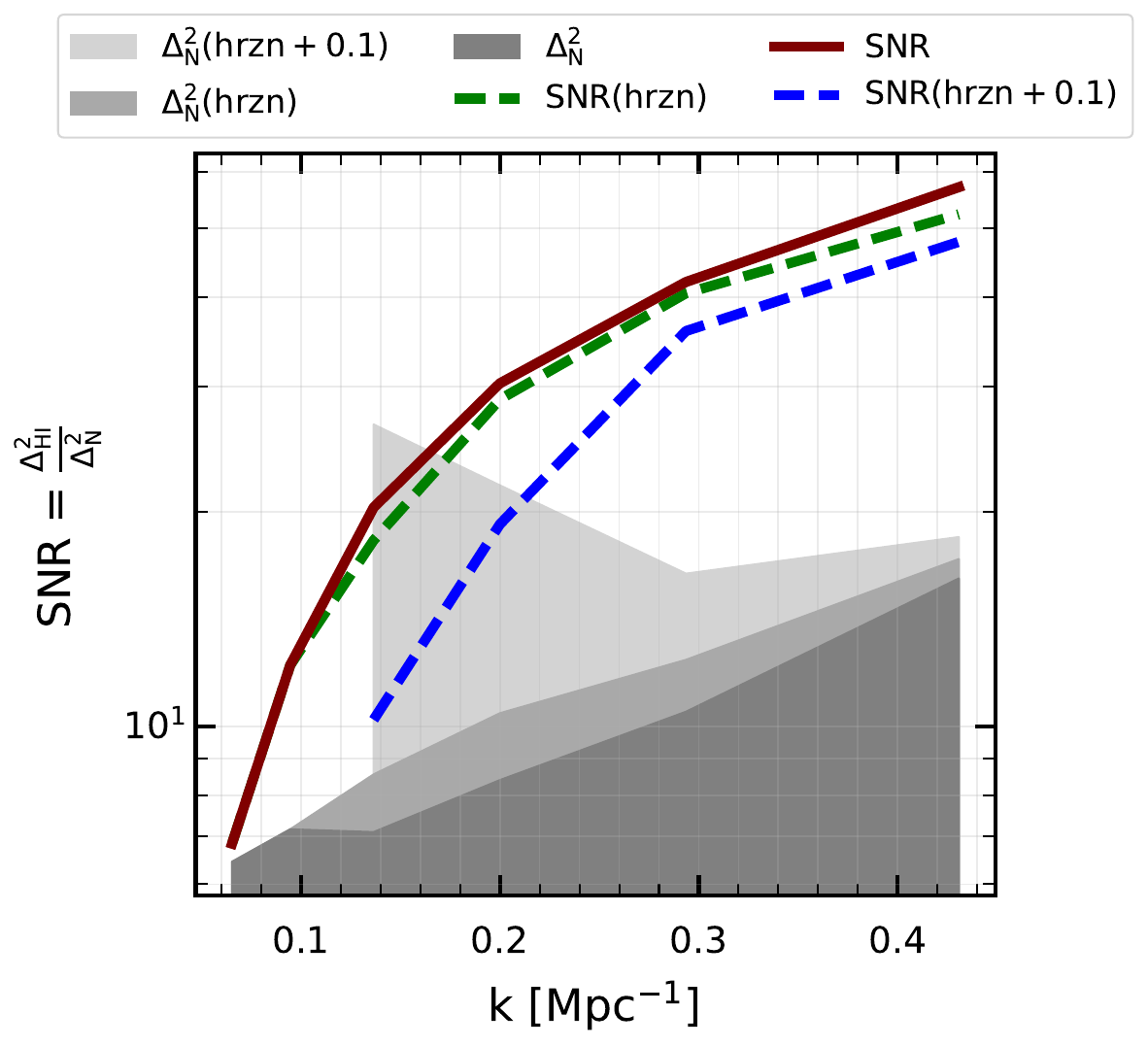}
  \caption{SNR plot for various cases, defined as the ratio of 1-D \HI\ power spectrum $\Delta^2_\text{HI}(k)$ to associated ``noise'' $\Delta^2_N(k)$. The noise is the per $k$-bin combined uncertainty due to thermal noise and sample variance, as indicated by vertical error bars in 1-D power spectra (Figure \ref{fig_4:gain_err_1d_ps_comp}). The brackets ``hrzn'' and ``hrzn + 0.1'' (dotted lines) indicate that foreground avoidance was applied about the horizon and horizon plus 0.1 Mpc$^{-1}$ buffer, respectively. Shaded regions represent total sensitivity ($\Delta^2_\text{th}(k) + \Delta^2_\text{sv}(k)$) for no avoidance (dark grey), avoidance about the horizon (medium grey), and the horizon plus buffer (light grey).}
  \label{fig_5:SNR_plot}
\end{figure}

Figure~\ref{fig_5:SNR_plot} illustrates the sensitivity curves and resulting SNR for each case. The second scenario (horizon-only avoidance) yields sensitivities nearly identical to the first, deviating only at small scales where thermal noise begins to dominate over sample variance. The average SNR loss across all $k$-modes is approximately 5\%. However, in the third scenario, the aggressive avoidance strategy masks a significant portion of the $k$-space, limiting the usable range to $k > 0.1$ Mpc$^{-1}$ and resulting in a substantial average SNR loss of $\sim$30\%.

In practice, real observations are often affected by more severe foreground leakage into the EoR window. Thus, careful tuning of the avoidance boundary is crucial to ensure that the retained $k$-modes are dominated by noise rather than systematics. A typical diagnostic technique is to plot 2D log power spectra from cross-correlations of time-interleaved visibilities, identifying EoR-like regions based on the presence of uniformly distributed positive and negative pixels \cite{MWA_UL_2016_FG_avoid}. Alternatively, methods such as filtering \cite{hybrid_paper_2018} or subtraction techniques (explored in the following sections) can be used to clean foreground-contaminated modes within the wedge, effectively improving SNR by expanding the usable region of the EoR window.

\subsection{Principal Component Analysis (PCA)}
PCA is one of the simplest blind foreground subtraction techniques in 21-cm cosmology \cite{blind_fore_sub_alonso}. While PCA is classically a linear dimensionality reduction method, it can be repurposed for foreground removal by exploiting the spectral smoothness and frequency correlations of foregrounds.
Assuming a noiseless model, our data consist of:
\begin{equation} 
\vec{d}(\nu,\vec{u}) = \vec{d}_{fg}(\nu,\vec{u}) + \vec{d}_{21}(\nu,\vec{u}),
\label{eq:d_dfg_d21}
\end{equation}
where $\vec{d}$ is the observed data, $\vec{d}_{fg}$ the foreground component, and $\vec{d}_{21}$ the cosmological signal, all as functions of frequency $\nu$ and baseline vector $\vec{u}$.

The PCA steps are as follows:
\begin{itemize}
    \item Mean-subtract the data along the LoS: $\vec{d}(\nu,\vec{u}) \rightarrow \vec{d}(\nu,\vec{u}) - \langle\vec{d}(\nu,\vec{u})\rangle_{\nu}$
    \item Compute the covariance matrix: $\boldsymbol{C} = \frac{\vec{d}(\nu,\vec{u})\cdot\vec{d}^{\dagger}(\nu,\vec{u})}{N-1}$
    \item Diagonalise $\boldsymbol{C}$ using SVD: $\boldsymbol{\hat{U}^{T}C\hat{U}} = \boldsymbol{\hat{\Lambda}}$, with $\lambda_i > \lambda_{i+1}$ for all $i$
    \item Choose $N_{fg}$ principal components that capture most of the foreground variance; form matrix $\boldsymbol{W}$ using the corresponding eigenvectors and project onto the data to form the foreground model: $\vec{f} = \boldsymbol{W}^{\dagger}\vec{d}$
    \item Subtract the model: $\vec{r} = \vec{d} - \vec{f}$ and use $\vec{r}$ for power spectrum estimation
\end{itemize}

We choose $N_{fg}$ such that the signal at the largest scale ($k \sim 0.05$ Mpc$^{-1}$) is not suppressed. Although PCA effectively removes low-rank foreground modes, it struggles to model non-smooth components introduced by mode-mixing or polarisation leakage. More advanced treatments, such as transfer functions \cite{transfer_func_MEERKAT_2023}, can address signal loss but are not pursued here.

\subsection{Gaussian Process Regression (GPR)}
A Gaussian Process (GP) is defined as a collection of random variables such that any finite subset follows a joint Gaussian distribution \cite{GPR_for_ML}. In GPR, we model the data using an infinite-dimensional distribution, defined by a mean and a covariance (kernel) function.

GPR has emerged as a powerful foreground removal technique, both in simulations \cite{Soares_2021,mertens_gpr_18} and in real observations (LOFAR \cite{LOFAR_UL_2020,LOFAR_UL_2024,LOFAR_UL_2025}; HERA \cite{HERA_GPR}). Starting again from Eq.~\ref{eq:d_dfg_d21}, we define the total kernel:
\begin{equation}
    K = K_{21} + K_{FG},
\end{equation}
with $K_{FG} = K_{\text{sky}} + K_{\text{mix}}$ separating smooth foregrounds and instrument-induced mode-mixing. We construct a joint distribution for the observed data and the foreground model:
\begin{equation}
[\vec{d}, \vec{d}_{\text{fg}}] \sim \mathcal{N}\left( \mathbf{0}, 
\begin{bmatrix} 
K_{21}+K_{FG} & K_{FG} \\
K_{FG} & K_{FG} 
\end{bmatrix} \right).
\end{equation}
The foreground mean and covariance estimates are then:
\begin{equation}
\label{eq:exp_cov_fg_2}
\begin{aligned}
E[\vec{d}_{fg}] &= K_{fg}(K_{fg}+K_{21})^{-1}\vec{d}, \\
\text{cov}[\vec{d}_{fg}] &= K_{fg} - K_{fg}(K_{fg}+K_{21})^{-1}K_{fg}.
\end{aligned}
\end{equation}
Subtracting the predicted mean gives the residual:
\begin{equation}
r = \vec{d} - E[\vec{d}_{fg}].
\end{equation}
The covariance term provides the uncertainty in the subtraction, which can be used for bias correction. 

In practice, kernel components are inferred from the data by optimising hyperparameters. A popular and flexible choice is the Matérn kernel:
\[
K_{\text{Matern}}(\nu - \nu_0) = \sigma^2 \frac{2^{1 - \eta}}{\Gamma(\eta)} 
\left(\sqrt{2\eta} \frac{|\nu - \nu_0|}{\ell}\right)^\eta 
K_\eta\left(\sqrt{2\eta} \frac{|\nu - \nu_0|}{\ell}\right),
\]
where $\sigma^2$ is the variance, $\ell$ is the correlation lengthscale, and $\eta$ governs smoothness. Common choices include: RBF ($\eta \rightarrow \infty$), MATÉRN52 ($\eta=5/2$), MATÉRN32 ($\eta=3/2$), and exponential ($\eta=1/2$). For this work, we use RBF for sky foregrounds, Mat32/52 for mode-mixing, and exponential for the \HI\ signal. We note that although the standard GPR technique \cite{mertens_gpr_18} has the capability to model non-smooth foreground components, kernel selection remains a challenging task, especially in high-contamination scenarios \cite{aishrilla_chapman_23}. Hence, generic prior covariance functions like RBF or MAT32/52 may not always accurately represent the true covariances, which can lead to biases in the results in the low SNR regime \cite{kern_2021}. Although the newer renditions of GPR like VAE-GPR \cite{LOFAR_GPR_ML,LOFAR_UL_2024} use EoR simulations to train the kernels, this is beyond the scope of this work.
\subsubsection{Application}
\label{subsec:GPR_meth}
We apply PCA and GPR to the residual gridded visibilities after dirty image reconstruction, similar to \cite{mertens_gpr_18,LOFAR_UL_2020}. The data at this stage are one Fourier transform along the LoS away from the delay space. 

We use the \texttt{GPy}\footnote{\url{https://sheffieldml.github.io/GPy/}} package to construct kernels and maximise the log-marginal likelihood (LML):
\begin{equation}
\log p(\mathbf{d}|\nu,\theta) = -\frac{1}{2}\left(\mathbf{d}^{T}K^{-1}\mathbf{d} + \log|K| + n\log2\pi\right),
\end{equation}
where $\mathbf{d}$ is the data, $K$ the total kernel, $n$ the number of data points, and $\theta$ the hyperparameters.

Hyperparameter optimisation is performed using the \texttt{emcee} MCMC sampler. The best-fit values are used to construct the foreground model, which is subtracted from the visibilities. We verify the robustness of sampling by exploring different prior ranges.

Initial runs showed that with increasing gain error, the \HI\ coherence scale hyperparameter $l_{21}$ converged to the prior upper bound. Inspired by \cite{LOFAR_UL_2018}, we adopt a gamma prior with an expectation value of 0.1 MHz (the known \HI\ scale), corresponding to $\Gamma(0.5,5)$. In the Table~\ref{tab:gpr_hyperparapms}, we listed the hyperparameters used in GPR to subtract the residual foreground. The corner plots of the posterior probability distributions of GPR hyperparameters after MCMC sampling for different calibration errors are shown in Figure~\ref{fig_6:corner_plots}.

\begin{table}
    \begin{tabular}{c|c|ccc|ccc}
        \hline
        Parameters & $\eta$ & \multicolumn{3}{c|}{Prior} & \multicolumn{3}{c}{Posterior} \\
        \hline
        & & 0.1\% & 1\% & 10\% & 0.1\% & 1\% & 10\% \\
        \hline
        $\sigma_\text{sky}/\sigma_n$ & $\infty$ &  $\mathbb{U}(0.1,5)$ & $\mathbb{U}(0.1,1)$ & $\mathbb{U}(0.1,5)$ & 1.02 & 1.03 & 0.48 \\
        $l_\text{sky}$ (MHz) &  &  $\mathbb{U}(5,50)$&$\mathbb{U}(5,50)$ & $\mathbb{U}(1,25)$ &46.62 & 24.11 & 3.03 \\
        \hline
        $\sigma_{21}/\sigma_n$ & 1/2 & $\mathbb{U}(10^{-6},10^{-1})$ &$\mathbb{U}(10^{-6},10^{-1})$ &  $\mathbb{U}(10^{-8},10^{-1})$ &2$\times$$10^{-3}$&  $10^{-3}$ & $0$  \\
        $l_{21}$ (MHz) &  & $\Gamma(0.5,5)$&$\Gamma(0.5,5)$  & $\Gamma(0.5,5)$  & 0.51 & 26.39 & 11.53 \\
        \hline
    \end{tabular}
    \caption{The GPR kernel priors and posteriors for 0.1\%, 1\% and 10\% gain calibration error cases.}
    \label{tab:gpr_hyperparapms}
\end{table}

\begin{figure}
    \centering
    \includegraphics[width=.55\textwidth]{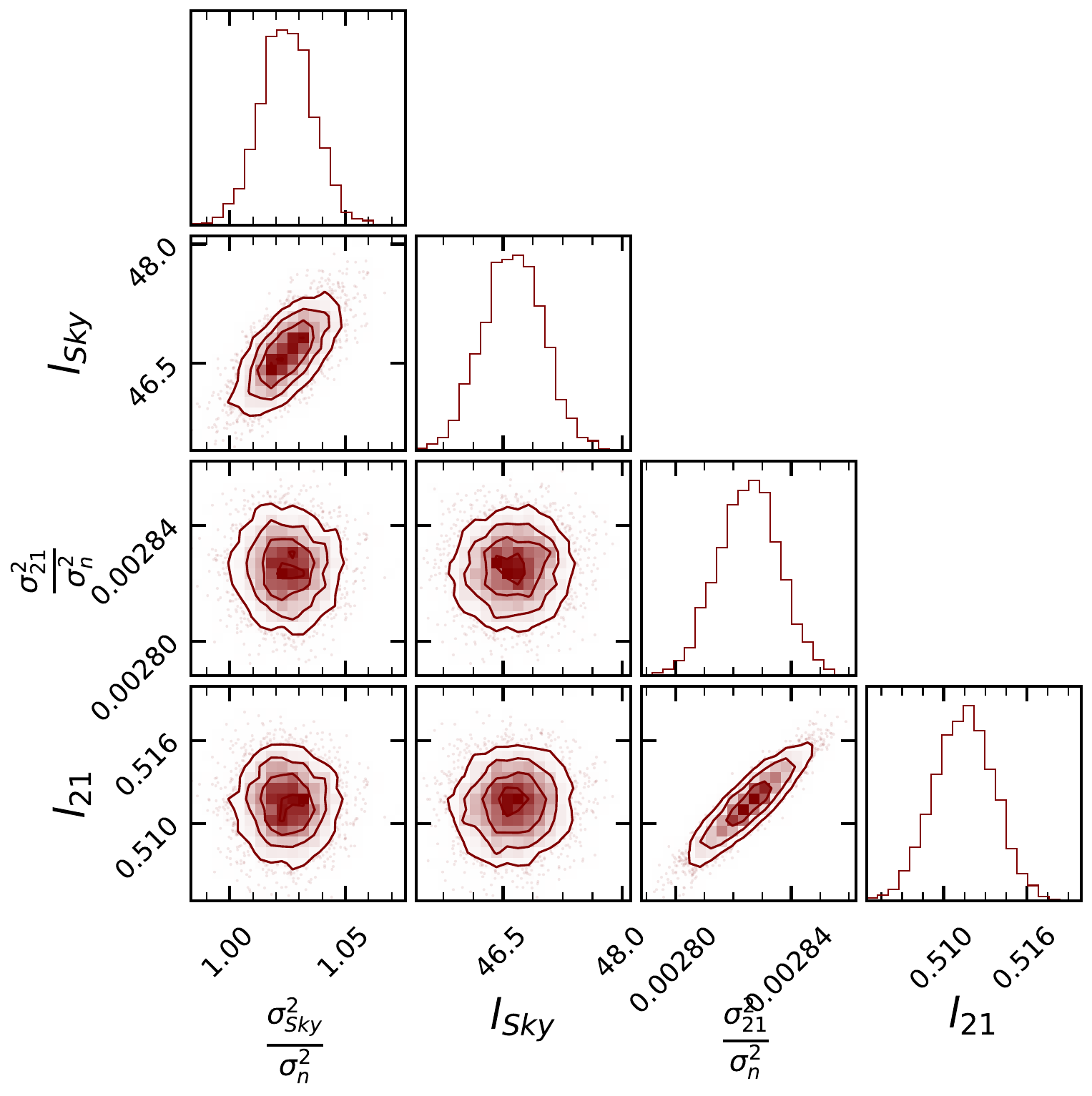}
    \caption{Posterior distributions of GPR hyperparameters after MCMC sampling for 0.1\% gain calibration error.}
    \label{fig_6:corner_plots}
\end{figure}

\section{Results}
\label{sec:results}
In this section, we discuss the results obtained after application of three distinct residual foreground mitigation strategies: avoidance, subtraction, and a hybrid approach. The subtraction methods use statistical modelling of foregrounds using either PCA or GPR, while the avoidance technique excludes $k$-modes below a designated wedge boundary, thereby separating regions heavily contaminated by residual foregrounds. The hybrid approach sequentially combines both subtraction and avoidance strategies. We aim to evaluate the robustness of each technique based on its effectiveness in recovering the true underlying \HI\ signal. In the following subsections, we discuss how different foreground mitigation strategies recover the target signal.  

\subsection{Case I: Gain Error of 0.1\%}
\label{subsec:results_0.1}
In this case, we incorporate a gain calibration error of 0.1\% and corrupt the simulated observed visibilities accordingly. The impact of a 0.1\% gain calibration error on 1D PS is presented in Figure~\ref{fig_4:gain_err_1d_ps_comp}. In addition, the absolute fractional error of 2D PS, computed using Equation~\ref{eq:2dps_frac_err}, is shown in the top row of Figure~\ref{fig_8:gain_err_1_10_0.1_2d_frac_ps_comp}. The 1D PS indicates that on larger scales ($k \leq 0.2$ Mpc$^{-1}$), the residual power closely matches the target signal level, even without applying any mitigation. However, on smaller scales ($k > 0.2$ Mpc$^{-1}$), the residual power underestimates the true signal by nearly an order of magnitude.

The plots in the top row in Figure~\ref{fig_8:gain_err_1_10_0.1_2d_frac_ps_comp} show the variation in 2D absolute fractional error maps for 0.1\% calibration gain error, after PCA subtraction, and after GPR subtraction, respectively. One can see that the subtraction techniques do not affect the absolute fractional error significantly. The corresponding spherically averaged power spectrum after the subtraction techniques is presented in the left plot of the top-row, Figure~\ref{fig_9:gain_err_1_1d_ps_comp}.
Using foreground avoidance around the horizon line, we manage to bring down the residual power to within 2$\sigma$ levels of the true \HI. The hybrid approach with GPR/PCA  improves the signal recovery on the small scales (right plot of the top row, Figure~\ref{fig_9:gain_err_1_1d_ps_comp}).

\subsection{Case II: Gain Error of 1\% }
\label{subsec:results_1}

\begin{figure}
  \centering
    \includegraphics[width=.75\textwidth]{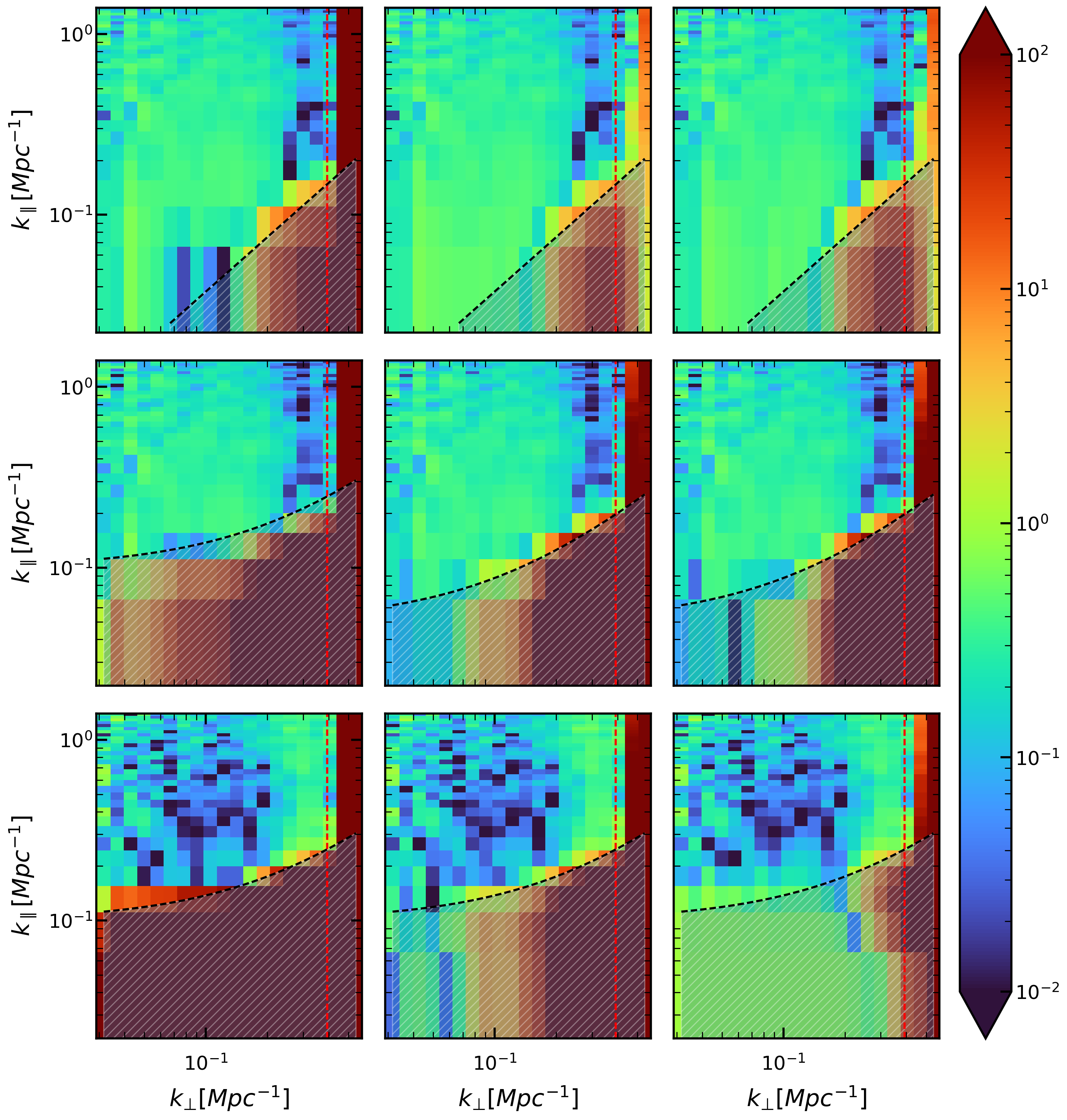}
    \caption{The 2D absolute fractional error maps $\Delta P_i(k_\perp, k_\parallel)$ for 0.1\% (\texttt{top row}), 1\% (\texttt{middle row}) and 10\% (\texttt{bottom row}) gain calibration errors. Columns (from left to right) correspond to sky-model subtraction, PCA subtraction, and GPR subtraction. The dotted black lines and the shaded regions within indicate the wedge boundaries and the masked $k$-modes used for avoidance for that particular case. The vertical red dotted line marks the scale at which PSF begins to dominate. Bluer regions indicate better agreement with the true \HI\ signal.}
    \label{fig_8:gain_err_1_10_0.1_2d_frac_ps_comp}
\end{figure}
To further quantify the impact of calibration systematics, we introduce a gain calibration error of 1\% and corrupt the simulated visibilities accordingly. The resulting cylindrical power spectrum is shown in the middle row of Figure~\ref{fig_3:gain_err_2d_ps_com}, while the corresponding spherically averaged power spectrum is presented in Figure~\ref{fig_4:gain_err_1d_ps_comp}. A comparison between the residual power spectrum, affected by a 1\% gain calibration error, and the true \HI\ power spectrum reveals substantial discrepancies across all $k$ modes, with residual power consistently exceeding the true signal by a significant margin. 

\begin{figure}
  \centering
   \includegraphics[width=.44\linewidth]{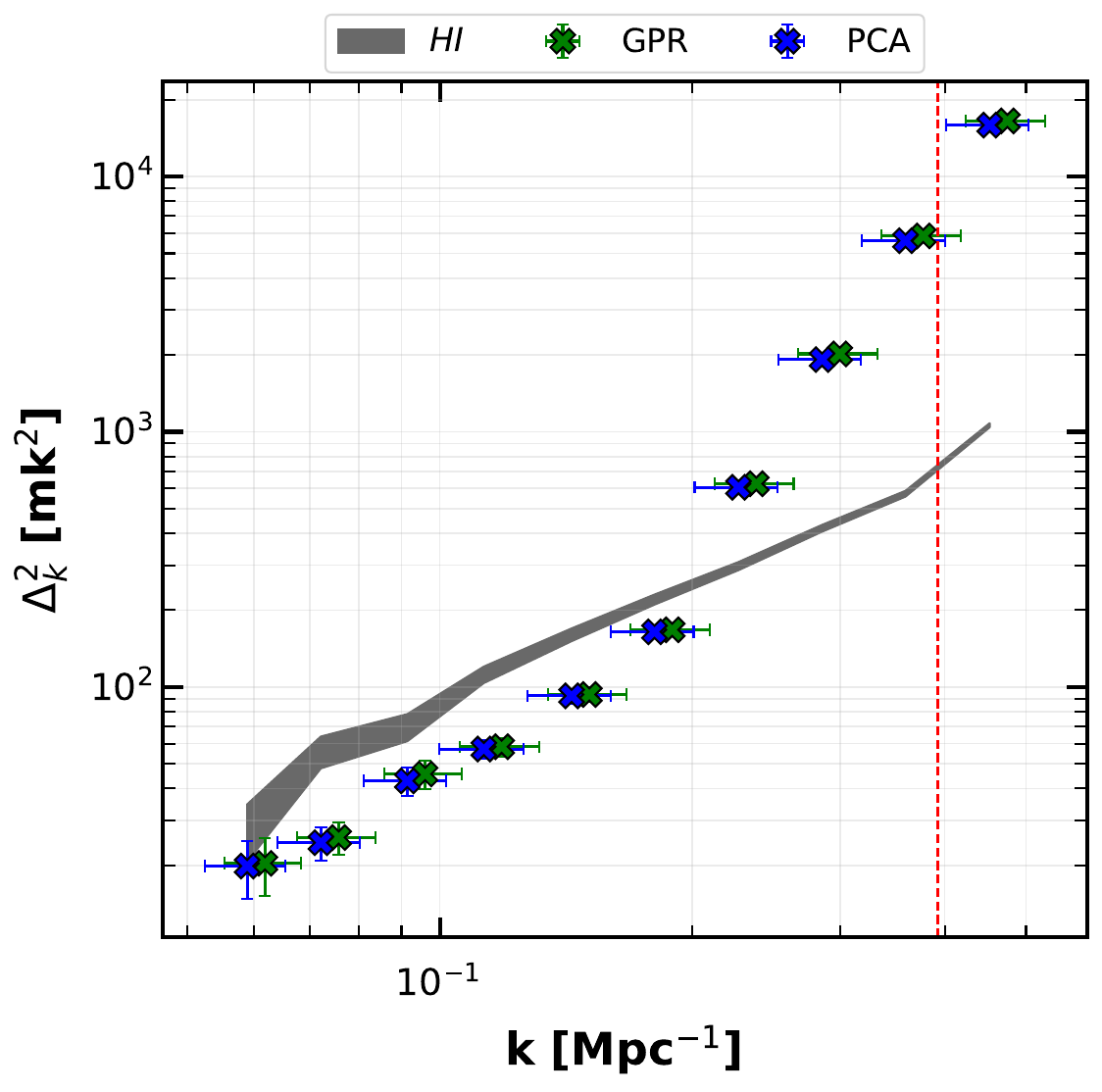}
    \quad
    \includegraphics[width=.44\linewidth]{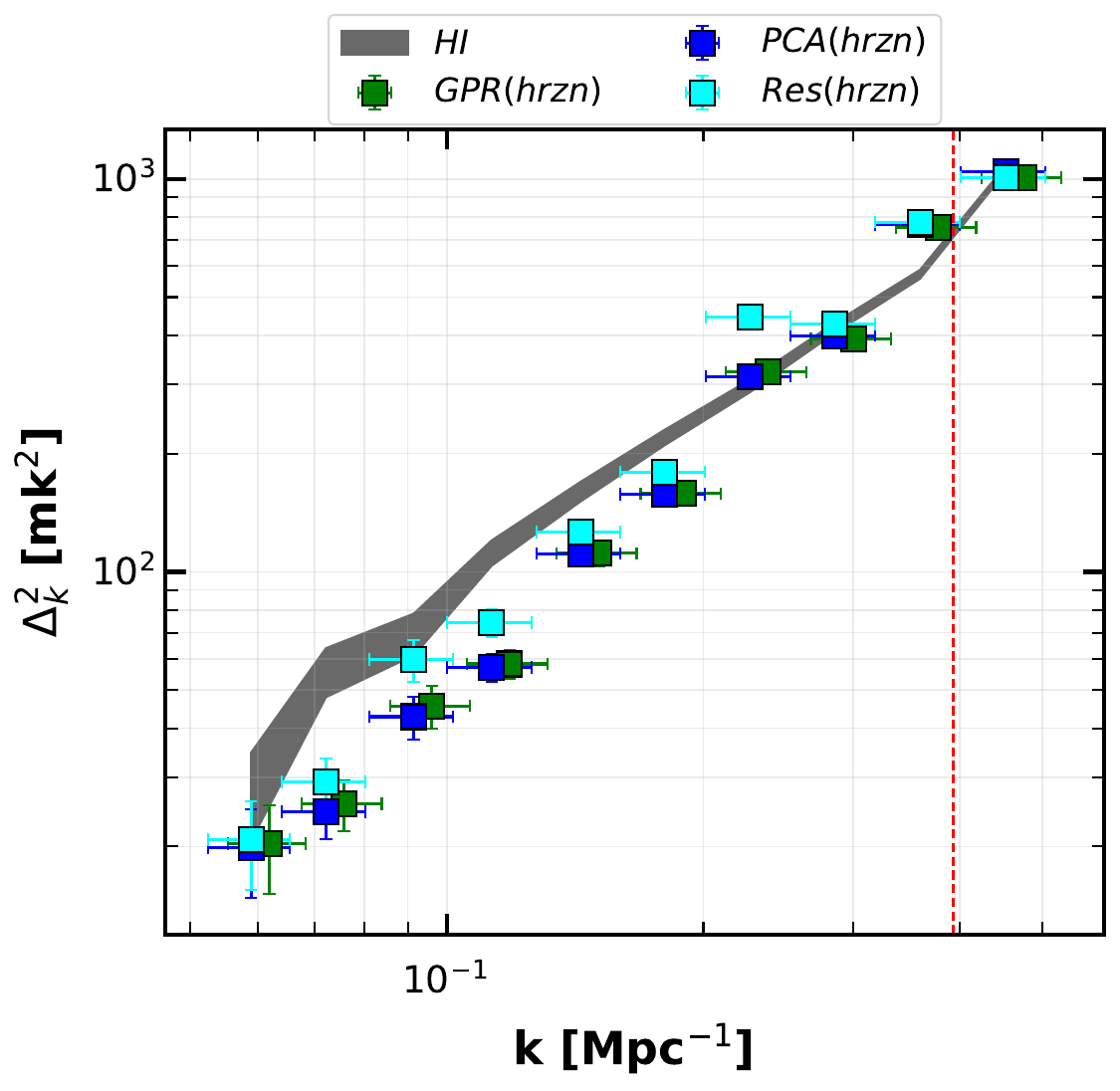}
    \quad
    \includegraphics[width=0.44\textwidth]{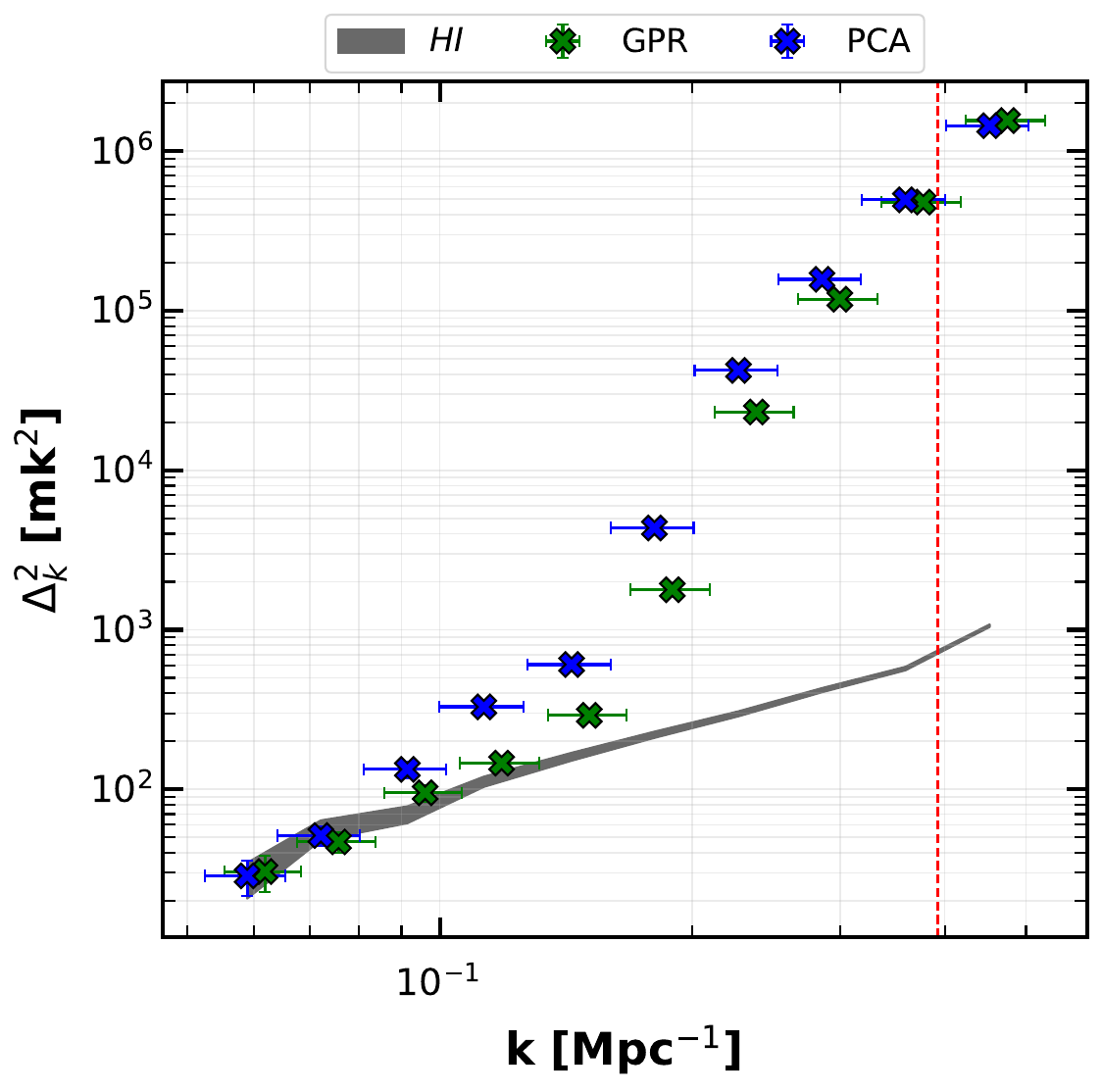}
    \quad
    \includegraphics[width=0.44\textwidth]{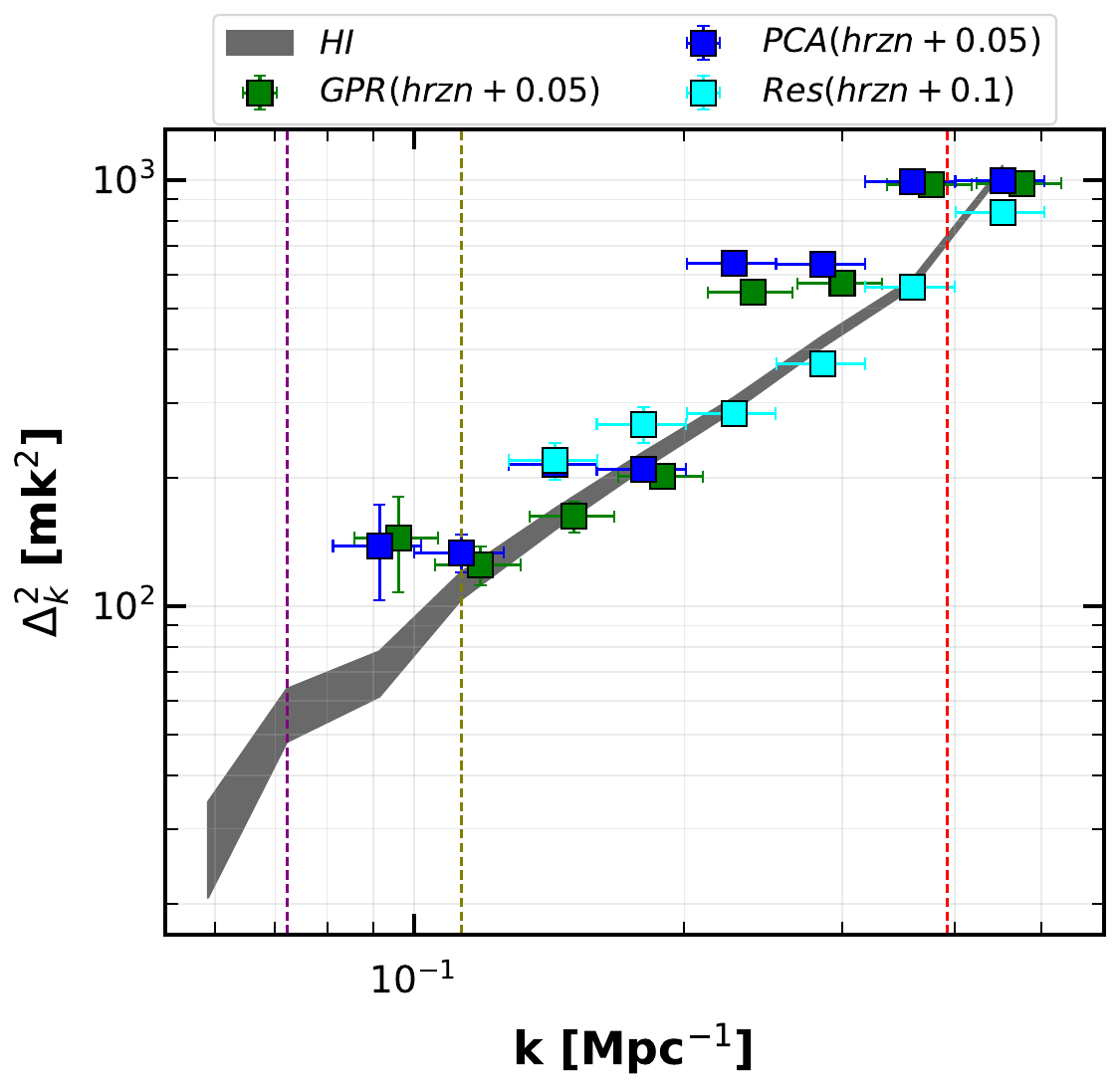}
    \quad
    \includegraphics[width=0.44\textwidth]{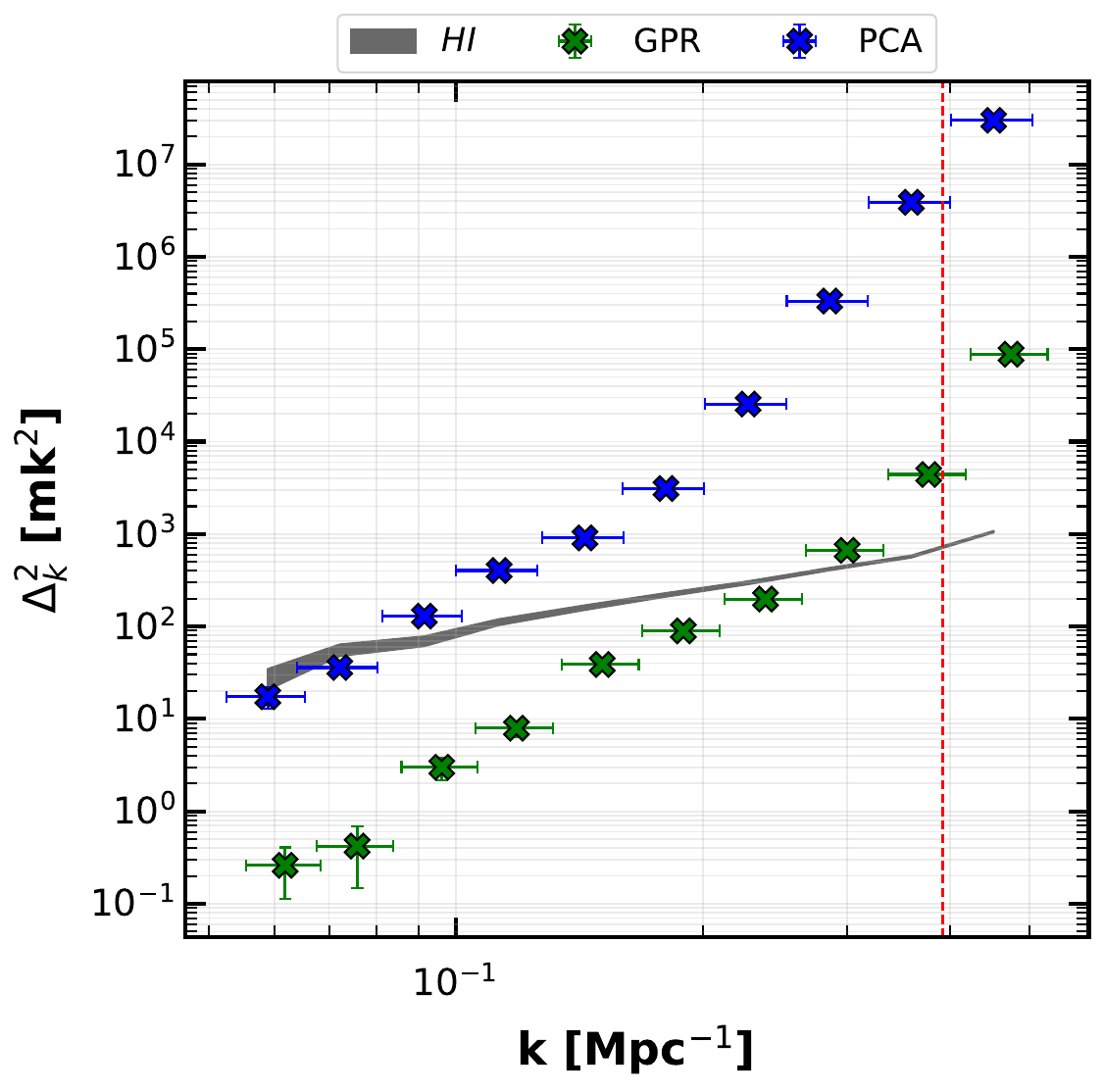}
    \quad
    \includegraphics[width=0.44\textwidth]{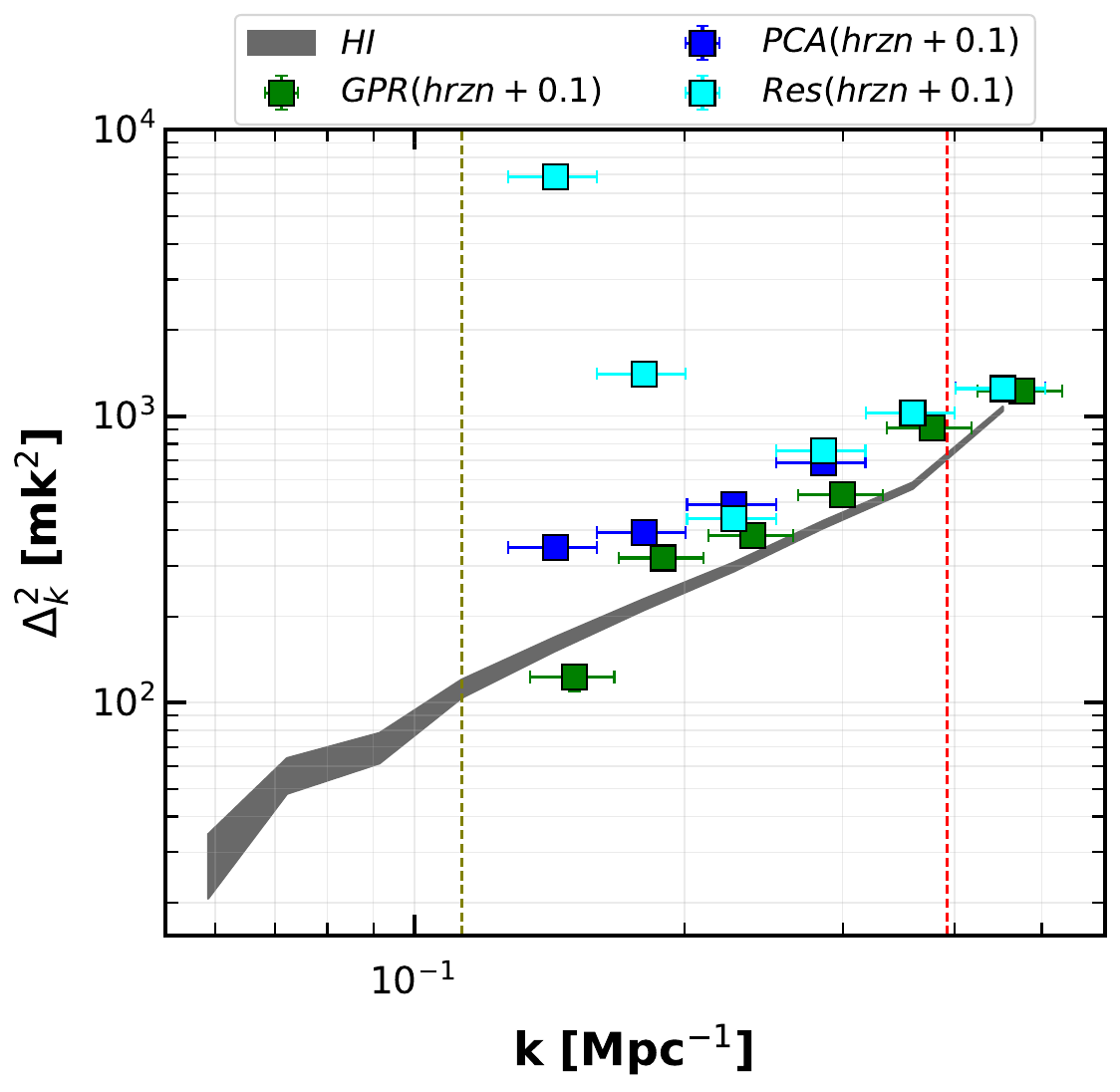}
    \caption{Spherically averaged 1D PS after residual foreground mitigation for 0.1\% (\texttt{top-row}), 1\% (\texttt{middle-row}) and 10\% (\texttt{bottom-row}) gain calibration error. \texttt{Left (Subtraction only)}: Star markers denote PCA- and GPR-subtracted residuals. \texttt{Right (Hybrid)}: The cyan markers show avoidance about the respective wedge boundaries (indicated by purple and yellow dotted vertical lines); green and blue markers represent hybrid (subtraction + avoidance) cases. The shaded region represents the true \HI\ signal. Markers for GPR are offset by 5\% for clarity.}
    \label{fig_9:gain_err_1_1d_ps_comp}
\end{figure}

The plots in the middle row in Figure~\ref{fig_8:gain_err_1_10_0.1_2d_frac_ps_comp} show the variation in 2D absolute fractional error maps for 1\% calibration gain error, after PCA subtraction, and after GPR subtraction, respectively. In contrast to the 0.1\% error case, one can see that the subtraction techniques effectively reduce the absolute fractional error by up to an order of magnitude in the low $k_{\perp}$ range ($\leq 0.1$Mpc$^{-1}$). Although GPR performs marginally better for higher $k_{\perp}$ modes than PCA, the mode-mixing component dominates, and we do not observe a significant improvement with either method. The corresponding spherically averaged power spectrum after the subtraction techniques is presented in the left plot of the middle row, Figure~\ref{fig_9:gain_err_1_1d_ps_comp}.

Using avoidance around the horizon line or a 0.05 Mpc$^{-1}$ buffer, our model fails to recover the signal on large scales. However, avoidance with a buffer of 0.1 Mpc$^{-1}$ (Figure~\ref{fig_8:gain_err_1_10_0.1_2d_frac_ps_comp}, \texttt{middle-left}) we manage to bring down the residual power to within 2$\sigma$ levels of the true \HI\ but with an average SNR loss of at least 30\% (see Section~\ref{subsec:FG_avoid}), as indicated by the cyan markers in the right plot of the middle-row, Figure~\ref{fig_9:gain_err_1_1d_ps_comp}.\\
Lastly, using a hybrid technique of applying avoidance (with a 0.05 Mpc$^{-1}$ buffer) to PCA/GPR residuals, it enables us to bring down residual power within 2$\sigma$ levels of the true \HI\ in the available $k$-modes, as represented by the green and blue markers in the right plot of the middle-row, Figure~\ref{fig_9:gain_err_1_1d_ps_comp}.

\subsection{Case III: Gain Error of 10\%}
In this case, we corrupted the observed visibilities with a 10\% gain calibration error to assess its impact on the recovered power spectrum. The effects are illustrated in the cylindrical power spectrum (\texttt{bottom} row of Figure~\ref{fig_3:gain_err_2d_ps_com}) and the spherically averaged power spectrum (Figure~\ref{fig_4:gain_err_1d_ps_comp}). We further examined the effectiveness of subtraction-based mitigation techniques in this high-error scenario. The plots in the bottom row in Figure~\ref{fig_8:gain_err_1_10_0.1_2d_frac_ps_comp} show the variation in the 2D absolute fractional error maps for 10\% calibration gain error and after using subtraction methods. Applying GPR and PCA independently shows that subtraction alone is insufficient. GPR grossly overestimates the signal, with strong suppression on larger scales, while PCA fails to account for the strong mode-mixing component on smaller scales (left plot in the bottom-row of Figure~\ref{fig_8:gain_err_1_10_0.1_2d_frac_ps_comp}). Hence, neither method adequately captures the foreground covariance at this level of contamination. 
Foreground avoidance with a 0.1 Mpc$^{-1}$ buffer also fails to reduce the residual power to the signal levels for $k \leq 0.2$ Mpc$^{-1}$. However, 1D PS results in the right plot bottom-row of Figure ~\ref{fig_9:gain_err_1_1d_ps_comp} shows that combining avoidance with a 0.1 Mpc$^{-1}$ buffer along with GPR subtraction, though still inaccessible for $k < 0.1$ Mpc$^{-1}$, perform better on smaller scales compared to hybrid PCA ($n_{\text{fg}} = 3$).\\

\begin{figure}[!ht]
    \centering
    \includegraphics[width=.45\textwidth]{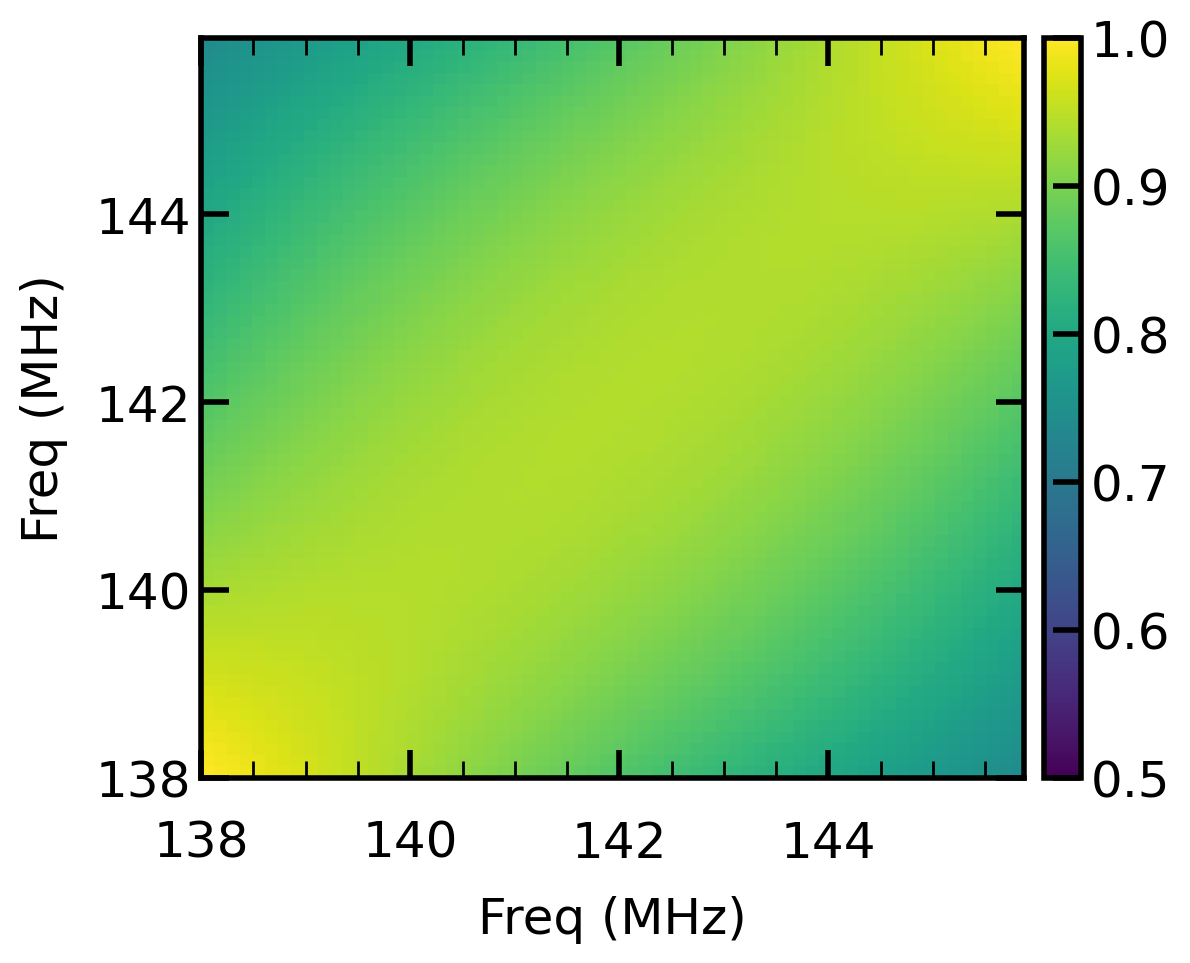}
    \quad
    \includegraphics[width=.45\textwidth]{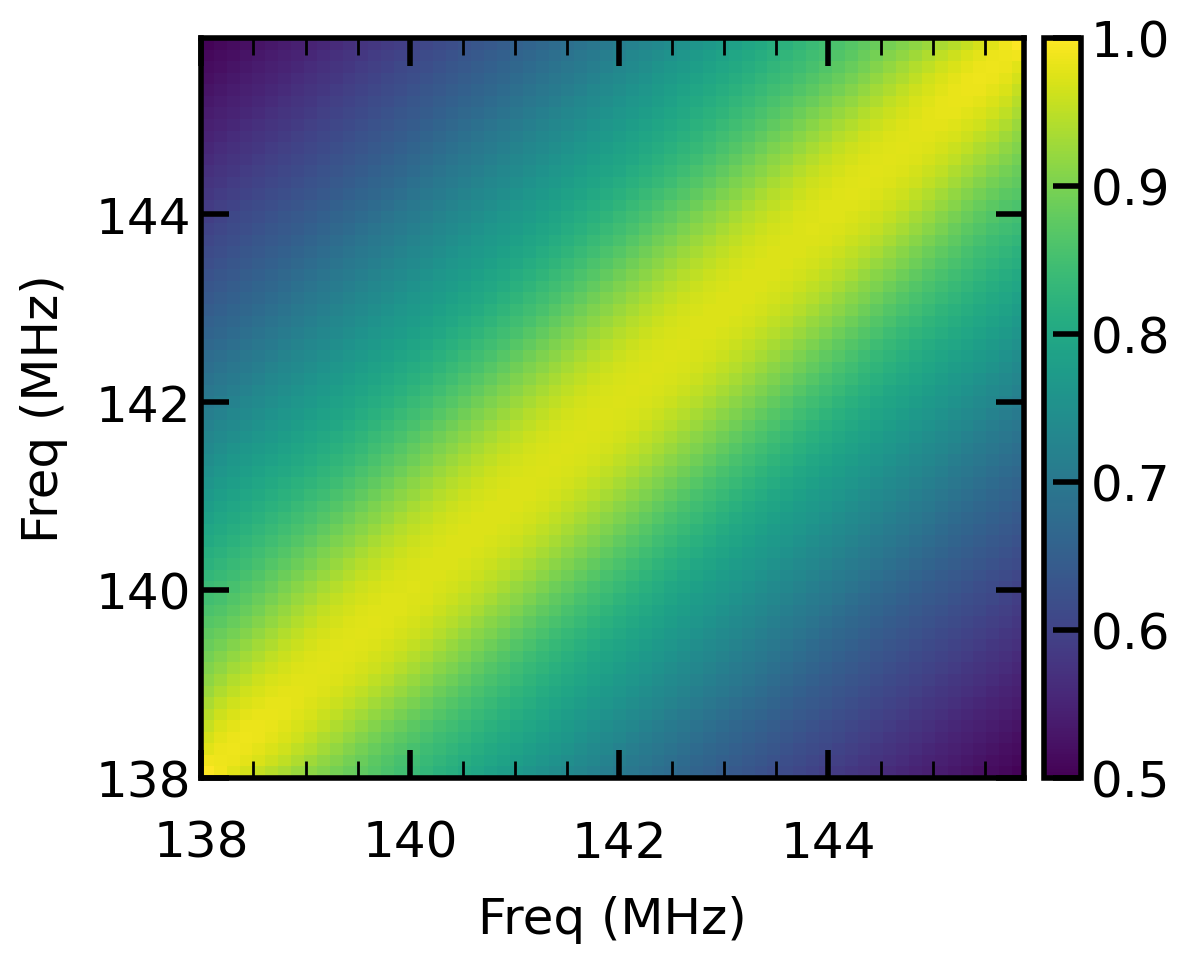}
    \caption{Frequency-frequency covariance of the GPR-reconstructed residual foregrounds for 1\% (\texttt{left}) and 10\% (\texttt{right}) gain errors without any mode-mixing components. Spectral smoothness degrades at higher gain errors, evident from diagonally auto-correlated structures in the 10\% case.}
    \label{fig_11:fg_cov}
\end{figure}

As can be seen in Figure~\ref{fig_11:fg_cov}, the frequency-frequency covariance map of the foreground model retrieved after GPR exhibits smoothly varying features across the bandwidth, reminiscent of foregrounds, for a 1\% gain calibration error, as opposed to comparatively less smooth features for a 10\% error. For both contamination levels, incorporating additional mode-mixing kernels (e.g., MAT-3/2 and 5/2) into the GPR model further suppressed large-scale power on smaller scales, while improving the Bayesian evidence and recovering $l_{21}$ more accurately (see Appendix~\ref{Appendix_1}). 

\section{Summary and Conclusion}
\label{sec:conclusion}
In this study, we examined the impact of gain calibration errors on statistical detection of the 21-cm signal and assessed the efficacy of foreground mitigation strategies on corrupted synthetic observations. We used the 21cmE2E pipeline to simulate the mock observation data for SKA1-Low AA* configuration. We computed 2D and 1D PS in the range of 0.05 $\leq$ k $\leq$ 0.5 Mpc$^{-1}$ from image cubes following three mitigation strategies: foreground subtraction (PCA, GPR), foreground avoidance, and a hybrid combination. It should be noted that we do not include galactic diffuse foregrounds, or frequency-dependent calibration gain errors in our simulations. Key findings for different gain error levels are summarised below:
\begin{itemize}
    \item Gain Error  $\sim 0.1$\%: Foreground avoidance about the horizon line is sufficient for recovery of the HI signal within 2$\sigma$ levels across all modes in the range 0.05 $\leq$ k $\leq$ 0.5 Mpc$^{-1}$. Subtraction or hybrid methods are redundant in this low-contamination regime.

    \item Gain Error $\sim 1$\%: GPR outperforms PCA on larger scales ($k \leq 0.1$ Mpc$^{-1}$), but both methods fall short on smaller scales. Foreground avoidance with a 0.1 Mpc$^{-1}$ buffer, or hybrid mitigation with a 0.05 Mpc$^{-1}$ buffer post-subtraction, we manage to constrain residual power within 2$\sigma$ of the true \HI\ across accessible $k$-modes till k $\leq$ 0.5 Mpc$^{-1}$. Avoidance with a 0.1 Mpc$^{-1}$ buffer results in an average SNR loss of approximately 30\%.

    \item Gain Error $\sim 10$\%: Subtraction (PCA or GPR) and avoidance (0.1 Mpc$^{-1}$ buffer) alone fail to recover the signal. GPR especially overestimates small-scale power while suppressing large-scale modes. However, hybrid mitigation using GPR combined with avoidance brings the residual power within 2$\sigma$ of the true \HI, in the range $0.1 \leq k \leq 0.5$ Mpc$^{-1}$. Again, this recovery comes at an average 30\% loss of SNR in the accessible $k$-modes.
\end{itemize}

Our results show that increasing gain errors contaminate the EoR window and erode the spectral smoothness of  extragalactic foregrounds, thereby limiting the effectiveness of standard subtraction-based methods. This raises a key question for future investigations: to what extent do  extragalactic foregrounds retain their smoothness when subjected to realistic instrumental effects and calibration systematics?  Our findings suggest that a hybrid approach, combining model-based subtraction with strategic avoidance, can recover the signal \HI\ even under moderate to severe contamination. However, this comes at the cost of reduced power spectrum sensitivity and signal-to-noise ratio on cosmologically relevant scales. Therefore, careful inspection of the wedge boundary and the spectral windowing becomes essential \cite{spectral_window_HERA_2023}.

Future work will extend this framework to account for additional observational systematics, including thermal noise, source position offsets, primary beam uncertainties, and bandpass calibration errors. We also plan to move toward Bayesian parameter estimation using emulator-based inference pipelines (e.g., \cite{tripathi_calibration}), and apply our methodology to real datasets, including upcoming Band-2 (120-250\,MHz) uGMRT observations at $z = 5.3, 6.3,\, 9.0\, \text{and}\, 10.2$ (Sagar et al., in preparation).

\acknowledgments
The authors thank anonymous reviewers for their valuable comments and constructive feedback. EB and AT thank the Indian Institute of Technology Indore for providing funding for this study in the form of a Teaching Assistantship. SKP acknowledges the financial support from the Department of Science and Technology, Government of India, through the INSPIRE Fellowship [IF200312]. The authors acknowledge the use of facilities procured through the funding via the Department of Science and Technology, Government of India, sponsored DST-FIST grant no.  SR/FST/PSII/2021/162 (C) awarded to DAASE, IIT Indore.\\
\textbf{Software}: We acknowledge the usage of PYTHON programming language
( \url{https://www.python.org/} ), numpy (\cite{Numpy}), scipy (\cite{Scipy_2020}), astropy (\cite{astropy}), emcee (\cite{emcee_2013}), and matplotlib (\cite{matplotlib}) in this work.
%Appendix
\appendix
\section{GPR with mode-mixing}
\label{Appendix_1}
\begin{figure}
\centering
\includegraphics[width=0.48\textwidth]{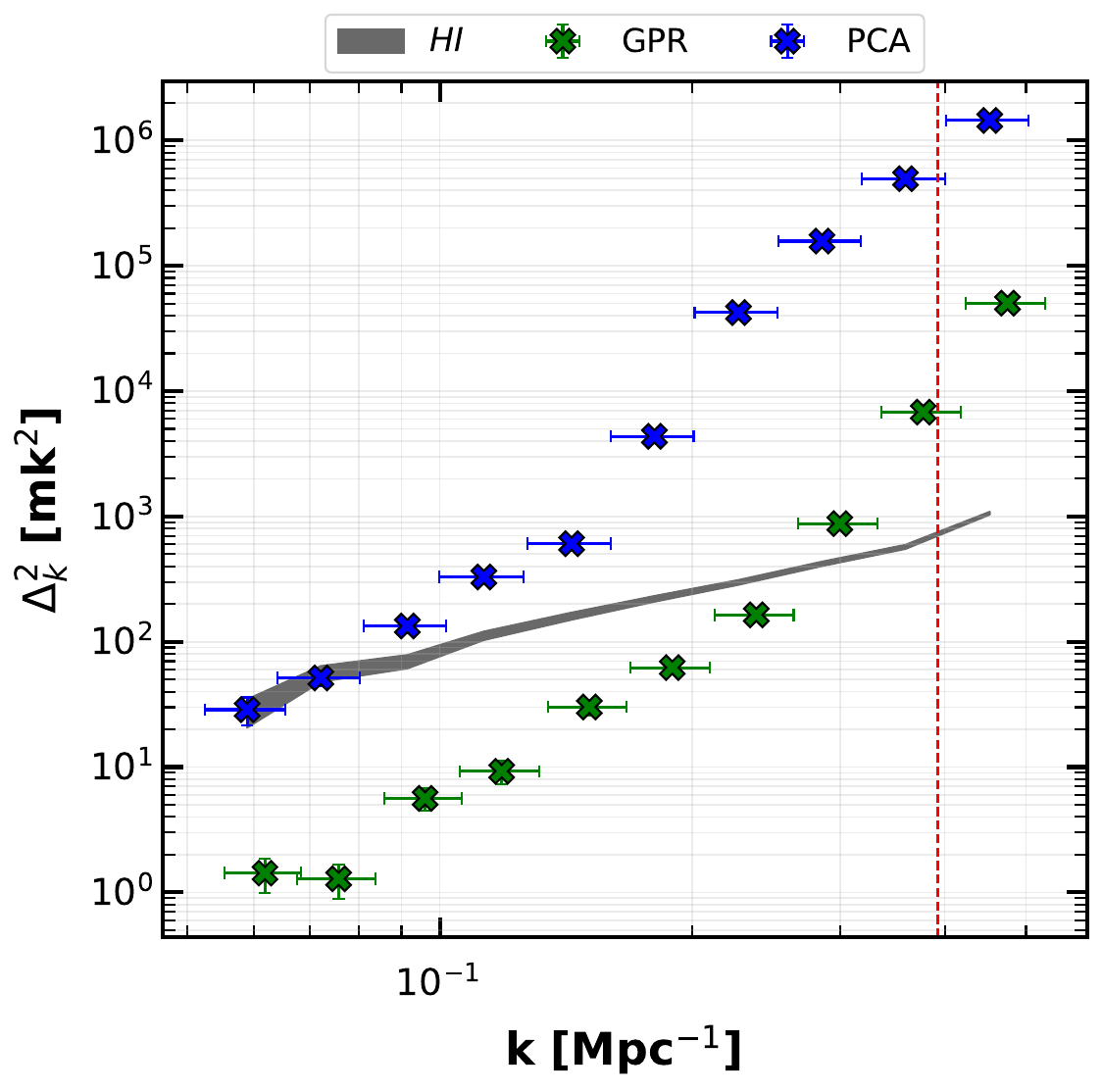}
\quad
\includegraphics[width=0.48\textwidth]{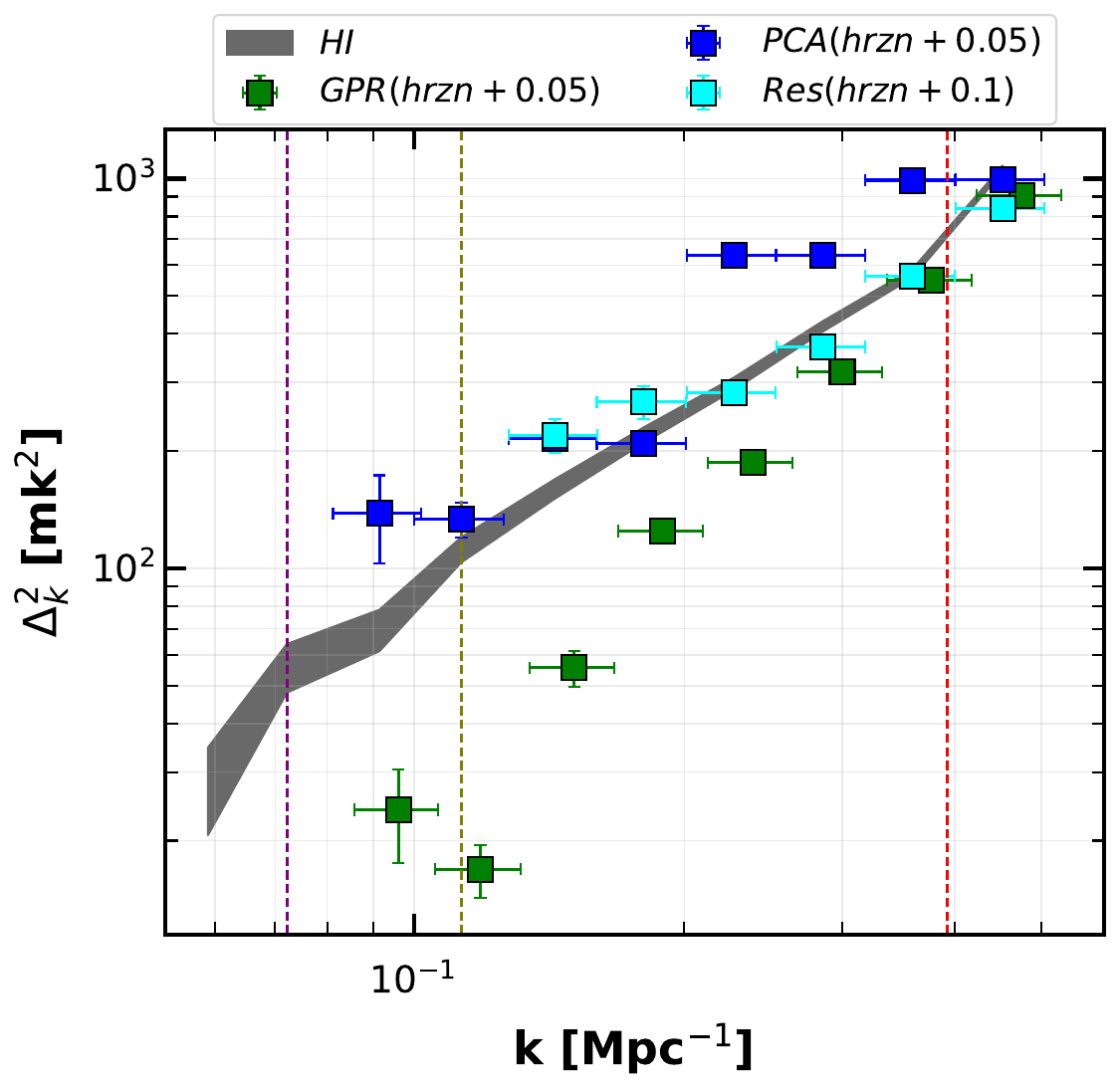}
\caption{Replication of Figure \ref{fig_9:gain_err_1_1d_ps_comp} but with mode-mixing effects included.}
\label{fig_12:gain_err_1_1d_ps_comp}
\end{figure}

\begin{figure}
\centering
\includegraphics[width=.48\textwidth]{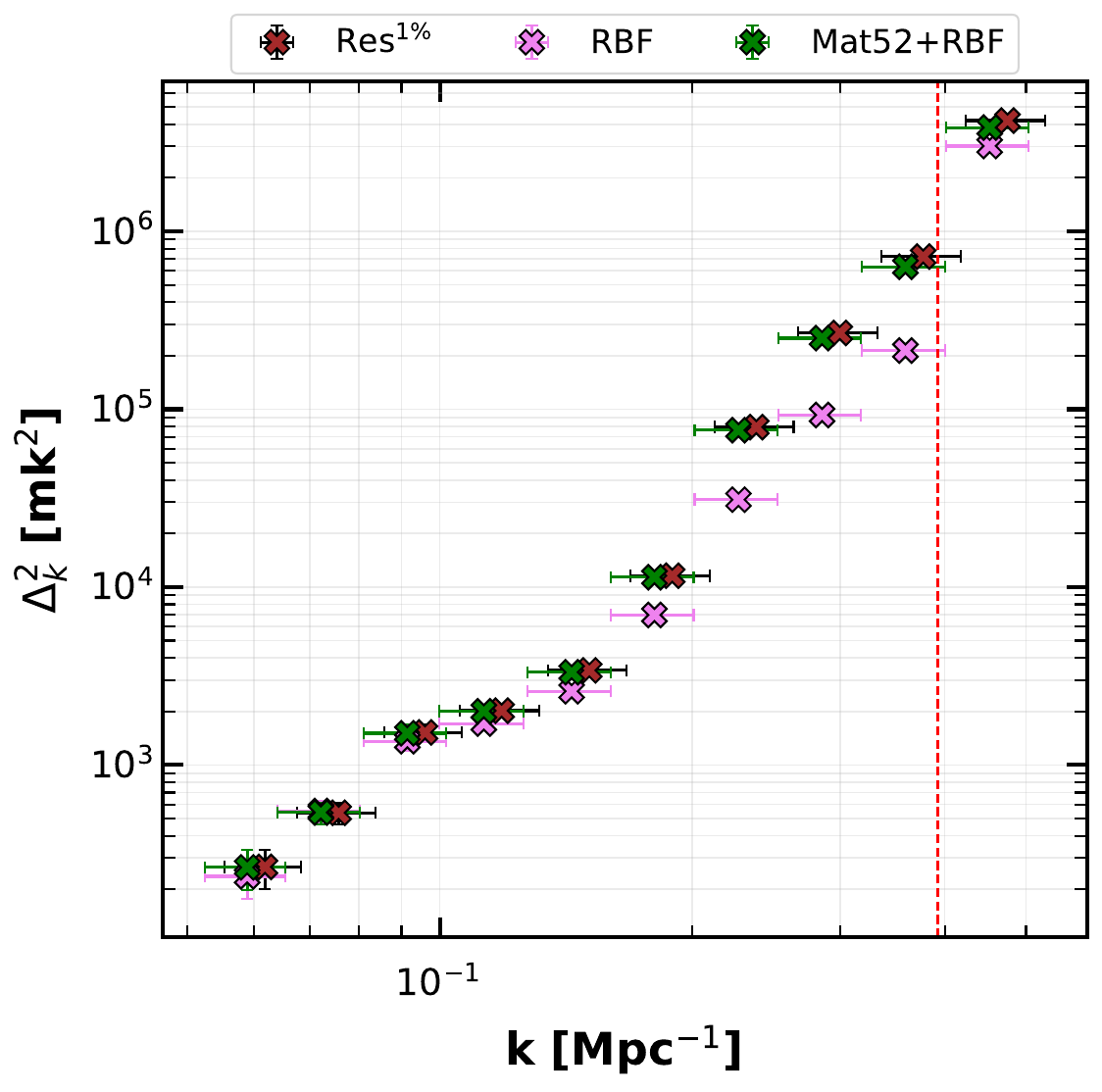}
\caption{The 1D PS of foreground model generated by GPR with (\texttt{green markers}) and without (\texttt{pink markers}) considering a mode-mixing prior. The \texttt{red markers} representing the power spectrum for residual foreground with 1\% gain calibration error are offset by 5\% along the $k$-axis for clarity.}
\label{fig_15:1d_ps_RBF_Mat52}
\end{figure}

\begin{figure}
\centering
\includegraphics[width=.75\textwidth]{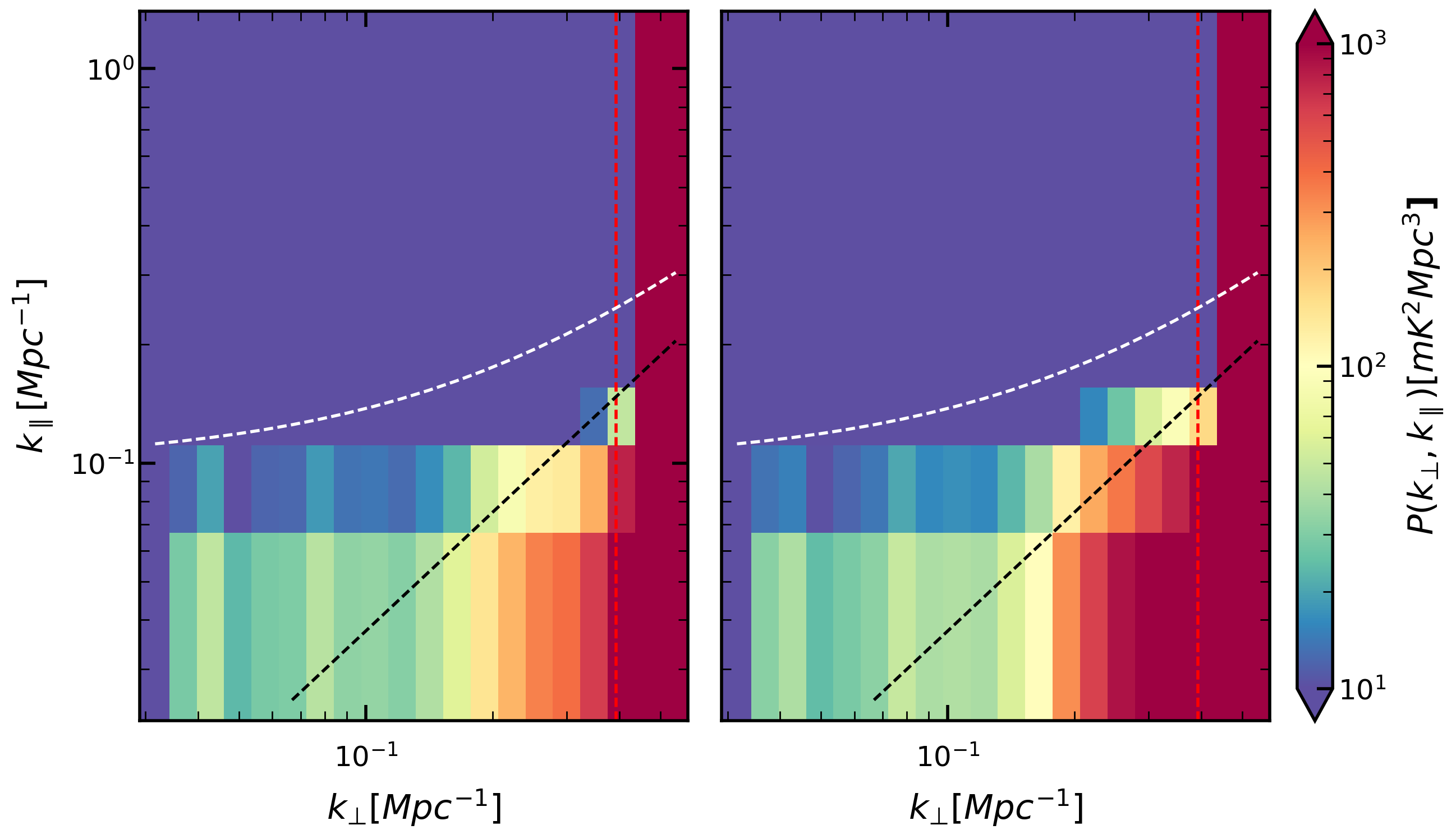}
\caption{The 2D PS of foreground model generated by GPR with (\texttt{right}) and without (\texttt{left}) considering a mode-mixing prior.}
\label{fig_14:2d_ps_RBF_Mat52}
\end{figure}

\begin{figure}
% \centering
\includegraphics[width=.75\textwidth]{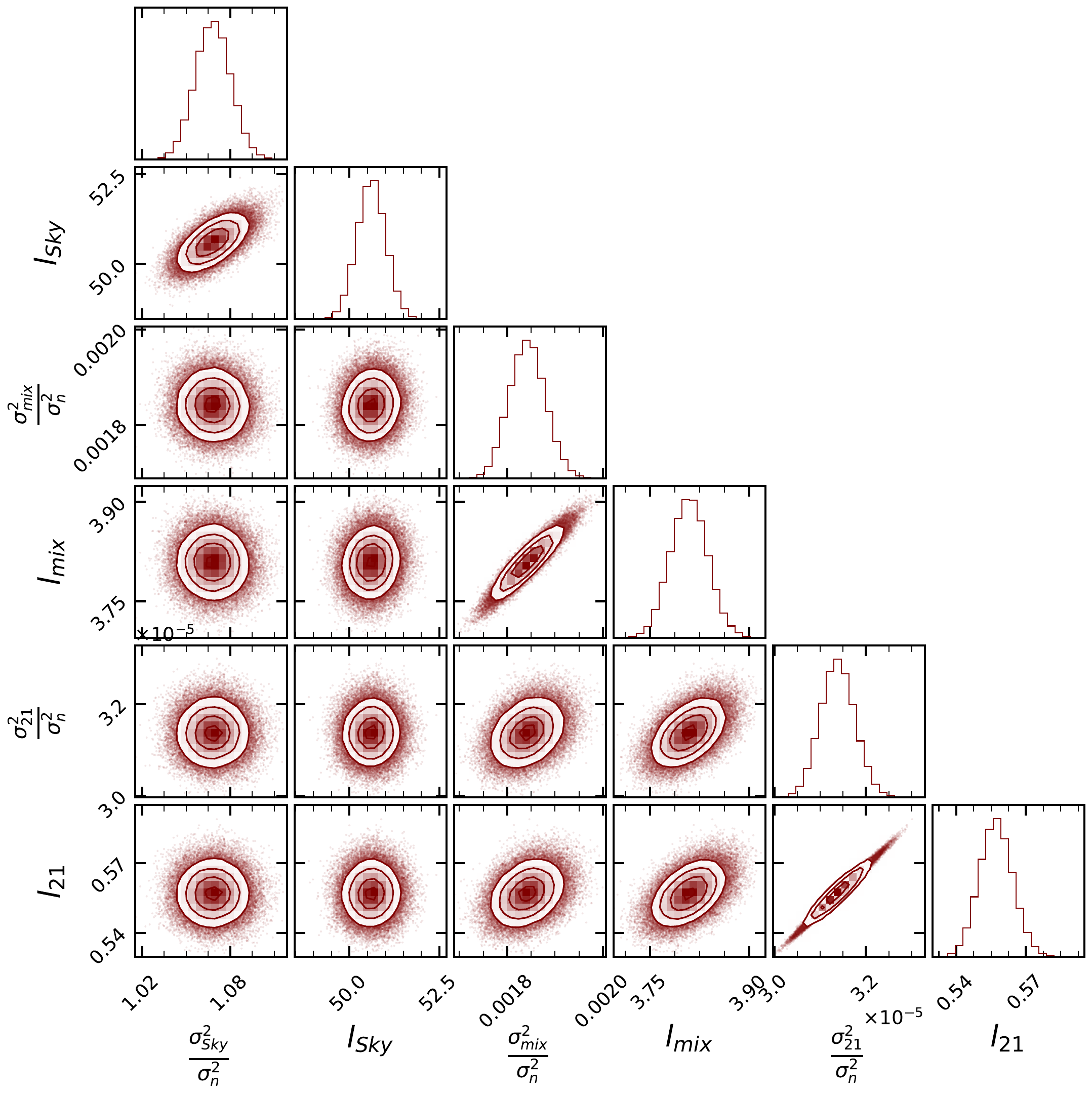}
\caption{Posterior distributions of GPR hyperparameters, including mode-mixing kernel for 1\% gain calibration error.}
\label{fig_13:corner_plot_gpr32}
\end{figure}

We present GPR results with a 1\% calibration error after accounting for mode-mixing kernel effects when generating the foreground model. As one can see from the 1D PS plot Figure \ref{fig_12:gain_err_1_1d_ps_comp}, there is over-subtraction on large scales (Figure \ref{fig_13:corner_plot_gpr32}).
The Figures \ref{fig_14:2d_ps_RBF_Mat52} and \ref{fig_15:1d_ps_RBF_Mat52} show the 2D PS and 1D PS of the GPR foreground models with (RBF + MAT52) and without (only RBF) a mode-mixing prior. The mode-mixing foreground model is picking up extra powers $k_{\perp}>0.1 Mpc^{-1}$, as compared to the non-mode-mixing model.

We present posteriors after MCMC for 1\% and 10\% calibration gain errors in the Figure \ref{fig_6:corner_plots}. Comparing with the posteriors obtained in the mode-mixing plus smooth sky foreground case (Figure \ref{fig_13:corner_plot_gpr32}), we see that $l_{21}$ is constrained closer to its true value when we take mode-mixing effects into account.
\begin{figure}
    \centering
    \includegraphics[width=.45\textwidth]{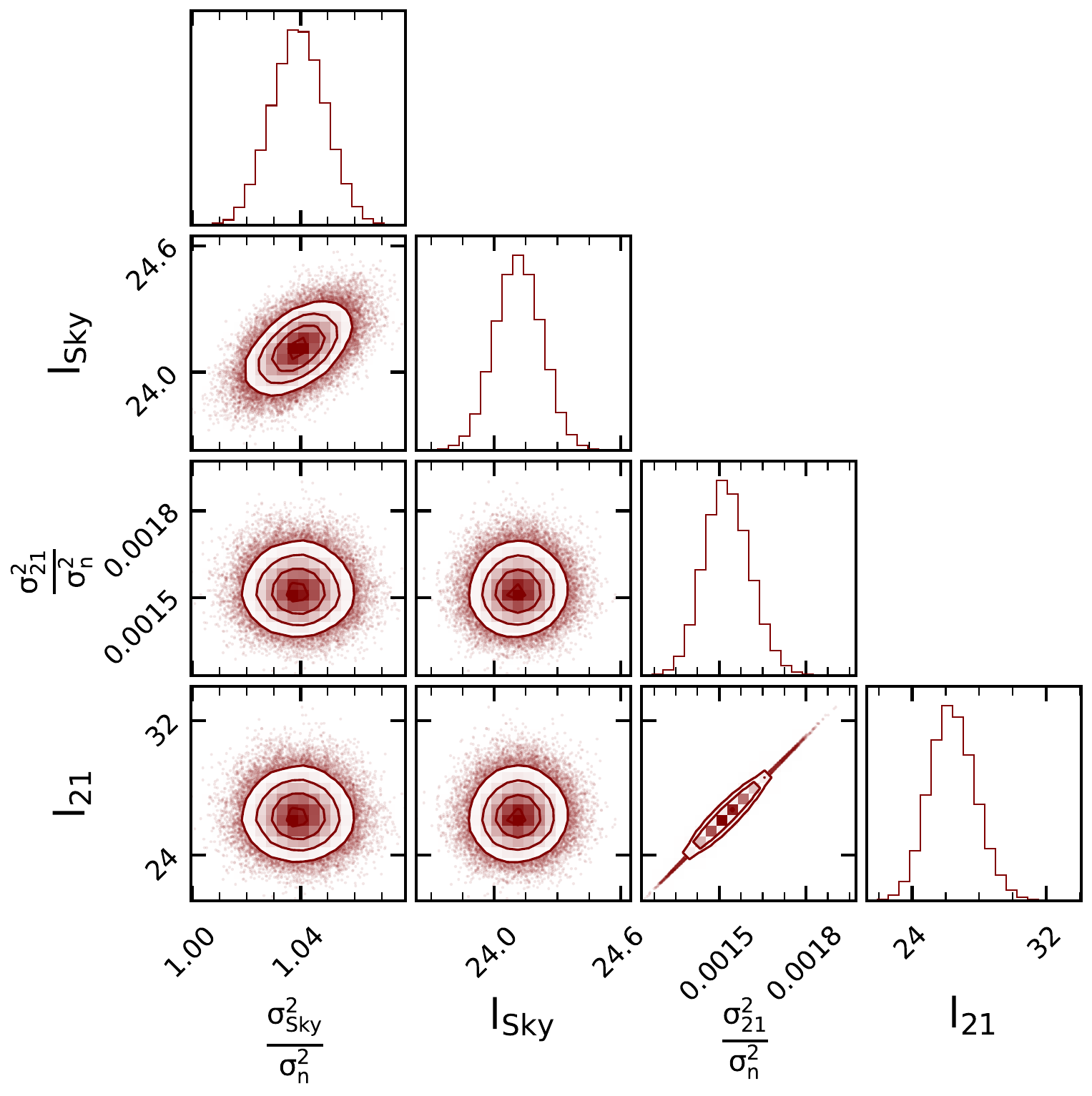}
    \quad
    \includegraphics[width=.45\textwidth]{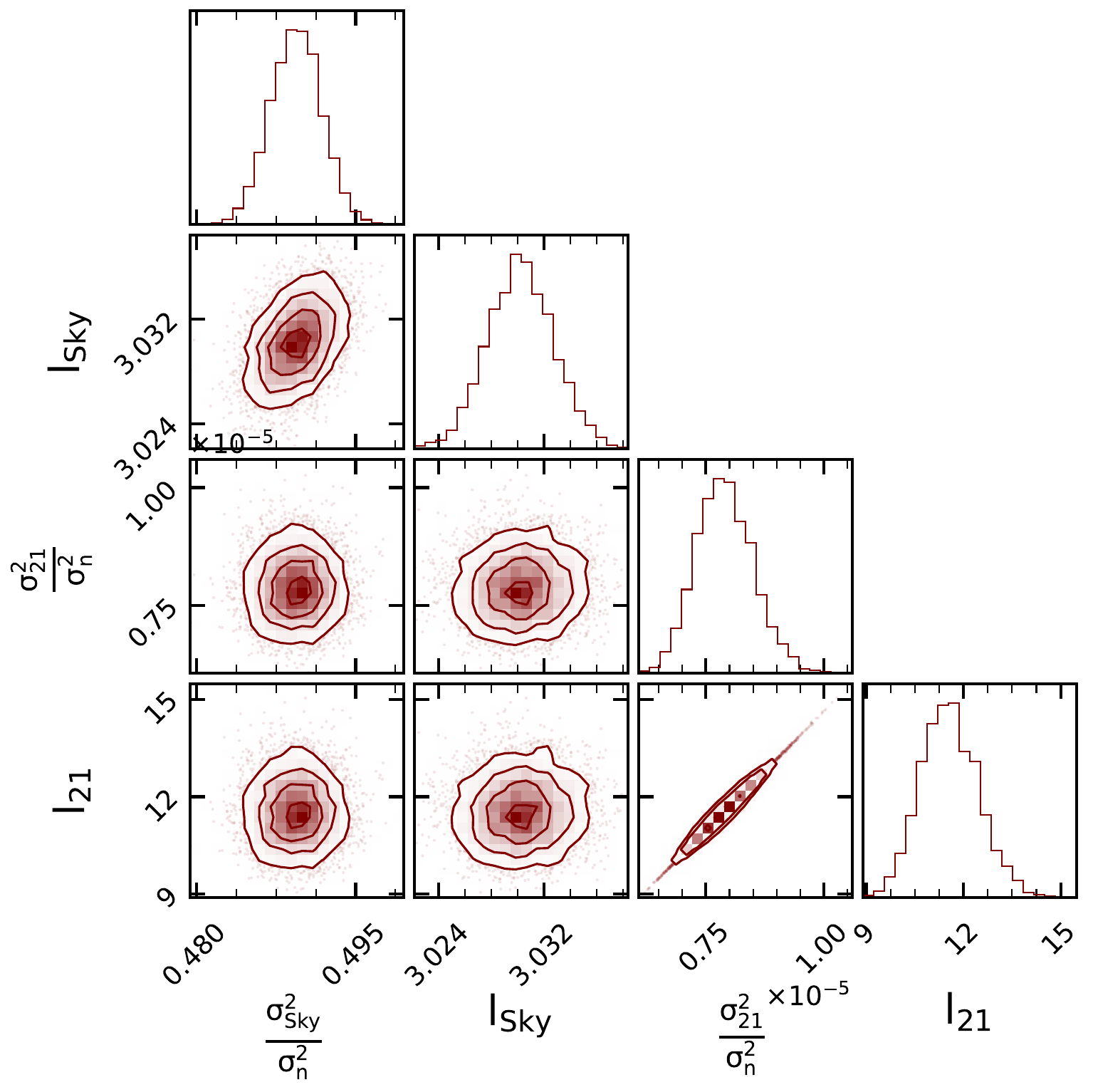}
    \caption{Posterior distributions of GPR hyperparameters without mode-mixing kernels after for 1\% (\texttt{left}) and 10\%(\texttt{right}) gain calibration errors. Contours represent confidence levels of 11.8\%, 39.3\%, 67.5\%, and 87.4\%.}
    \label{fig_6:corner_plots}
\end{figure}

% Bibliography
\bibliographystyle{JHEP}
\bibliography{biblio.bib}
\end{document}